\documentclass[useAMS,usenatbib]{mnras}

\usepackage{graphicx}
\usepackage[usenames,dvipsnames]{xcolor}
\usepackage{xspace}
\usepackage{float}
\usepackage{amsmath}
\usepackage{amsfonts}
\usepackage{amssymb}

\newcommand\HI{\ion{H}{i}}
\newcommand{\tir}{{\sc TiRiFiC}\xspace}
\newcommand{\new}[1]{#1}
\newcommand{\newtwo}[1]{#1}

\newcommand{\SoFiA}{{\sc SoFiA}\xspace}

\title[Non-parametric \HI morphometrics]{Non-parametric estimation of morphological lopsidedness}
\author[Giese et al.]
{Nadine Giese$^{1,2}$\thanks{E-mail: giese@astro.rug.nl},
Thijs van der Hulst$^1$,
Paolo Serra$^{3}$,
Tom Oosterloo$^{1,2}$
\\
$^{1}$University of Groningen, Kapteyn Astronomical Institute, Landleven 12, NL-9747 AD, Groningen, the Netherlands\\
$^{2}$ASTRON, the Netherlands Institute for Radio Astronomy, Postbus 2, NL-7990 AA Dwingeloo, the Netherlands\\
$^{3}$CSIRO Astronomy and Space Science, Australia Telescope National Facility, PO Box 76, Epping, NSW 1710, Australia
}
\begin{document}

\date{Accepted 2016 June 13. Received 2016 June 12; in original form 2015 December 05}     

\pagerange{\pageref{firstpage}--\pageref{lastpage}} \pubyear{2002}

\maketitle

\label{firstpage}
\begin{abstract}
  Asymmetries in the neutral hydrogen gas distribution and kinematics of galaxies are thought to be indicators \newtwo{for both} gas accretion and gas removal processes. These are of fundamental importance for galaxy formation and evolution. Upcoming large blind \HI surveys will provide tens of thousands of galaxies for a study of these asymmetries in a proper statistical way. Due to the large number of expected sources and the limited resolution of the majority of objects, detailed modelling is not feasible for most detections. We need fast, automatic and sensitive methods to classify these objects in an objective way. Existing non-parametric methods suffer from effects like the dependence on signal to noise, resolution and inclination. Here we show how to correctly take these effects into account and show ways to estimate the precision of the methods. We will use existing and modelled data to give an outlook on the performance expected for galaxies observed in the various sky surveys planned for e.g. WSRT/APERTIF and ASKAP.

\end{abstract}


\section{Introduction}
Hydrogen is the most abundant element in the Universe and neutral hydrogen (\HI) in particular is of utmost importance in the study of galaxy evolution. \HI can extend out to large radii beyond the optical disk, where it can efficiently trace episodes of gas accretion and removal. These include the inflow of gas from the intergalactic medium (IGM), the feedback from the galactic disk through e.g. star formation (``galactic fountain''), the ram pressure exerted by the gaseous medium within which galaxies move, as well as the in-fall of and interaction with gas-rich companions.
Observationally, these events can be revealed by the lopsidedness of the \HI morphology or kinematics, or by the warping of the \HI discs outer regions \citep{Sancisi2008}.

To understand how the above processes affect galaxy evolution we need to assemble large samples of \HI detected galaxies and explore the relation between \HI asymmetries and other galaxy properties. However, all large, blind \HI surveys carried out so far have inadequately large angular resolution for this purpose because they have been done using single dishes (e.g., HIPASS, ALFALFA and EBHIS; see \citealt{Barnes2001,Giovanelli2005,Kerp2011}). In the near future, radio telescopes with higher resolution, adequate sensitivity and high survey speed will for the first time enable this kind of investigations (e.g., the SKA pathfinders ASKAP and WSRT/APERTIF; see \citealt{Johnston2008,Verheijen2008,Oosterloo2010}).
The results of these future observing campaigns do not suffer as much from selection effects as previous, \new{smaller} surveys (e.g. WHISP, Atlas3D and THINGS; see \new{\citealt{vanderHulst2001,Serra2012I,Walter2008}}).

On average 1000 \HI detected objects per week are expected for WSRT/APERTIF. The large amount of acquired data can greatly contribute to the study of galaxy formation and evolution but at the same time bring about new challenges with regards to data reduction and analysis. One such challenge is the need to develop analysis methods that are fast, objective, reliable, and sensitive to physical processes that are significant for galaxy evolution.

Existing methods to determine \HI properties in galaxies focus on fitting models either to the full \HI data cube, the integrated \HI map, or the velocity field. 
For example, \tir \citep{Josza2007} and $^{3{D}}$BAROLO \citep{DiTeodoro2015} fit 3D tilted-ring models \citep{Warner1973,Rogstad1974} to data cubes. The possible fit parameters include surface brightness, rotational velocity, scale height, position angle, and inclination at different radii within the galaxy.
The rotation curve can also be obtained by fitting the velocity field using the task \emph{rotcur} \citep{Begeman1987} in the image processing software package {\sc gipsy} \citep{vanderHulst1992}. In both the 2D and the 3D cases, asymmetries could either be determined by comparing the true 3D distribution to a symmetrically fit model or by fitting the galaxies in separate parts, e.g. for the receding and approaching part of the disk and comparing parameters for those.  Another existing application for the detection of asymmetries is DiskFit \citep{Spekkens2007,Sellwood2010}, which can fit non-axisymmetric models to the velocity fields. DiskFit also offers the possibility to model asymmetries in photometric images.

With the large number of galaxies that are expected to be detected within ongoing and future large blind \HI surveys, detailed modelling or fitting of every single object will pose problems with respect to execution time, but also resolution of the \HI data, since the majority of objects will not have sufficient resolution for detailed fitting. The emphasis here lies on the objects with limited resolution, because only 0.05\% of all objects detected in a shallow all-sky survey with WSRT/APERTIF will have sufficient resolution to be fitted with \tir \citep[see][]{Duffy2012}. Methods using non-parametric approaches include determining asymmetries by harmonically decomposing the integrated \HI map \citep{vanEymeren2011} or determining indices adopted from characterisation in the optical \citep{Holwerda2011I,Holwerda2011II}.

In this paper, we re-examine the possibility of using the \new{morphometric} parameters employed in \cite{Holwerda2011II} for a non-parametric characterisation of \HI in galaxies. In section \ref{section:asymmetry} we will investigate the dependence of different \new{morphometric} parameters on data qualities such as noise and angular resolution as well as the inclination of objects.
In section \ref{section:bias} we show a possible way to correct the Asymmetry parameter for noise effects and in section \ref{section:conclusion} we summarise our findings and discuss future work.

\section{Non-parametric measurement of lopsidedness}\label{section:asymmetry}
  \subsection{Measuring morphological shape revisited}
    \cite{Holwerda2011II} found that using a suitable combination of parameters adopted from optical studies, one can distinguish \new{interacting} from \new{non-interacting} galaxies. They use a combination of indices measuring Concentration, Asymmetry, Smoothness, Gini, $M_{20}$ and a newly defined parameter Gini-$M_{20}$ \newtwo{(\textit{GM})} on a subsample of the WHISP survey that had been visually classified in \cite{Noordermeer2005} and \cite{Swaters2002}. We will define these indices below. Comparing the visual classifications with their automatically determined parameters, \cite{Holwerda2011II} propose to use either a combination of the Asymmetry parameter and $M_{20}$ or of the Concentration and $M_{20}$ to determine the boundary between interacting and non-interacting galaxies in parameter space. In this Section we revisit their results.

One aspect that deserves attention is the effect of a varying S/N. In particular, the \HI data used by \cite{Holwerda2011II} have relatively low S/N, and one may wonder whether this biases the measured parameter values. Consider, for example, the Asymmetry parameter (hereafter also referred to as \textit{A} or the \textit{A} parameter), which was first \new{used in \cite{Abraham1996} and} defined in \cite{Conselice2000} \new{(CBJ00)} as:

\begin{equation}
 A = \frac{\sum_{i,j}{\left|I(i,j)-I_{180}(i,j)\right|}}{2\sum_{i,j}{\left|{I(i,j)}\right|}},
 \label{eqn_asymmetry}
\end{equation}

\noindent where in the case of \HI data $I$ denotes the zero-th moment image of a galaxy cube. $I_{180}$  refers to the same image rotated by 180 degrees. 
\new{CBJ00 already show that for optical data the measured value of \textit{A} depends strongly on the S/N of the data even for S/N values of several hundreds. Furthermore, while A can be corrected within 5\% of the noise-free value for S/N$>100$, no reliable correction is possible below this limit. \cite{Lisker2008} investigated the effect of S/N and aperture size on the determination of the Gini parameter and also found poor performance at low S/N levels.
Since most \HI data will not reach a S/N level of 100, this aspect needs further attention. We discuss the influence of S/N on the results for \HI data in Sect.\ \ref{subsubsec:s2n}.
}

\new{These noise effects are also relevant for the study of} \cite{Holwerda2011II} who use the highest resolution \HI WHISP maps for their analysis. 
\new{The masks (3D segmentations of the cubes that contain signal from the galaxy) have been generated using all voxels above a 1.8 $\sigma$ threshold in the Hanning smoothed 60" data cube. The 2D \HI images were created as the zero-th moment of the masked version of the highest resolution data cube.}
To check the applicability of the \textit{A} parameter for 
data cubes of quality comparable to that of the WHISP ones, we estimated the mean and maximum S/N per galaxy of the full galaxy sample in an aperture containing 90\% of the galaxy flux. To this end we estimated the S/N along a line of sight using only the voxels in the mask:

\begin{equation}
(S/N)_{\textrm{i,j}} = \frac{\sum_{k}I_{i,j,k}}{\sqrt{N_{i,j}}\sigma}. \label{eq_s2n1}
\end{equation}

\begin{figure}
  \begin{center}
    \includegraphics[width=0.35\textwidth]{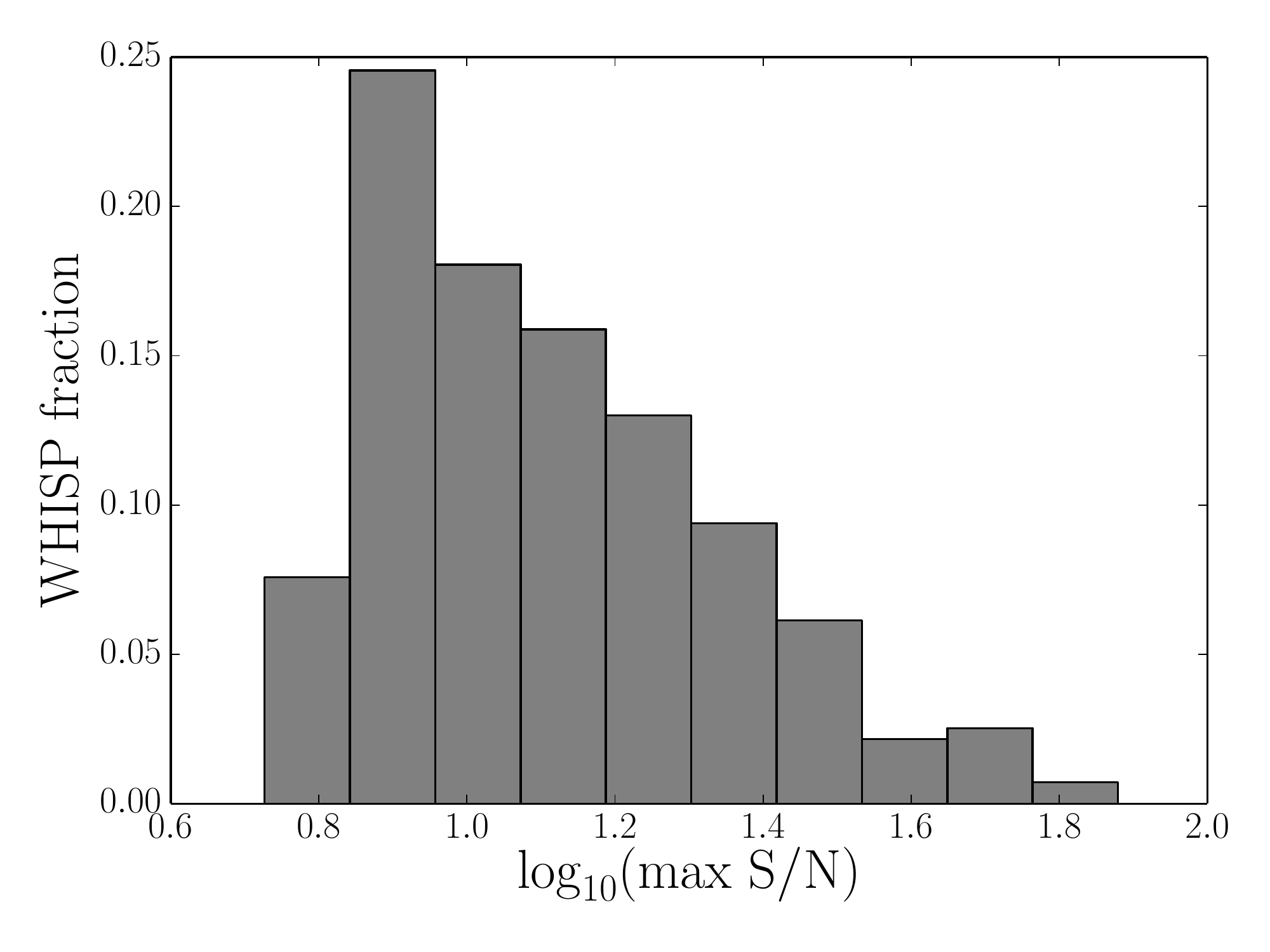}
    \includegraphics[width=0.35\textwidth]{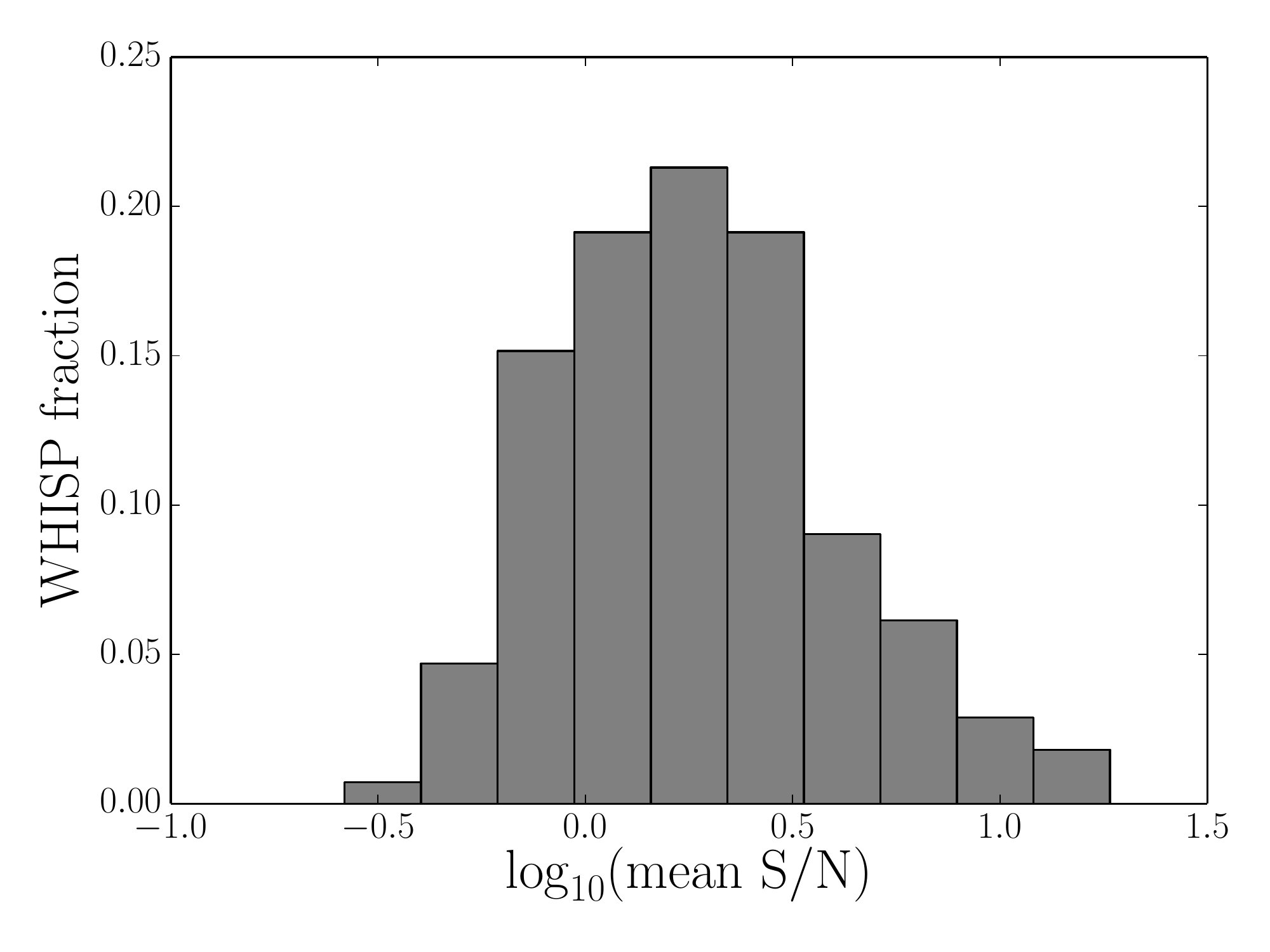}
  \end{center}
  \caption{Histograms of the mean and maximum pixel value in the WHISP high resolution signal to noise maps - For the majority of objects the mean signal to noise in the moment maps is below 10. The maximum signal to noise is below 100 in all cases.}
  \label{hist_s2n}
\end{figure}

Here, $I_{i,j,k}$ are the voxels that are included in the object mask, with the index $k$ referring to the velocity/frequency axis. $N_{i,j}$ is the number of channels that have been added to the pixel with coordinates (i,j) in the integrated \HI image. $\sigma$ is the RMS noise in the cube. We determine the global S/N for a galaxy as the average of the line of sight S/N distribution:
\begin{equation}
S/N = \langle (S/N)_{\textrm{i,j}}\rangle. \label{eq_s2n2}
\end{equation}
We find that the average as well as the maximum signal to noise is always significantly below the level of 100. Fig.\ \ref{hist_s2n} shows the distribution of the mean and maximum S/N values throughout the sample. Therein, the majority of galaxies have a mean S/N below 10 and the maximum pixel values are below 
100 for all galaxies. Thus, none of the objects have sufficient signal to noise to use the Asymmetry parameter without suffering from noise effects. In \cite{Holwerda2011II} no attempt was made to correct the \textit{A} values for noise. 

We used the same WHISP maps of the 141 galaxies analysed in \cite{Holwerda2011II} to measure the parameters mentioned above. \new{We chose the optical centres provided by \cite{Noordermeer2005} and \cite{Swaters2002} for the calculation of Concentration, Asymmetry, $M_{20}$ and  \textit{GM} . We excluded the 10 galaxies, for which \cite{Holwerda2011II} did not calculate any \new{morphometric} parameters. Additionally, we excluded UGC 10448 since the \HI observation has been carried out for the wrong target \citep[see][]{Noordermeer2006}. This left us with 130 objects. We used a $1\sigma$ threshold to select the pixels that are used for calculation. This threshold is low enough to include the outer disk of the sample galaxies. 
\cite{Holwerda2011II}, too, apply a threshold to the WHISP \HI images in order to measure the morphometric parameters. Their adopted threshold is, however, in \HI column density (the actual value is not given but we note that \cite{Holwerda2011I} adopt a threshold of $3\times10^{19}\mathrm{cm}^{-2}$ when analysing the THINGS sample). We prefer to use a S/N threshold because noise is the main driver of systematics in the measurement of these parameters. The actual choice of S/N threshold is not trivial, because it can have a significant effect on the calculated parameters. If a galaxy breaks up into several pieces because of a too high threshold or because the S/N is too low, the Asymmetry parameter will reach very high values close to the maximum, although a visual inspection might reveal that the object is symmetric. On the other hand, if the threshold is chosen too low, the Asymmetry value will reach higher values as well, due to the strong effects that noise has on pixels with low flux.}

\begin{figure}
  \begin{center}
    \includegraphics[width=0.45\textwidth]{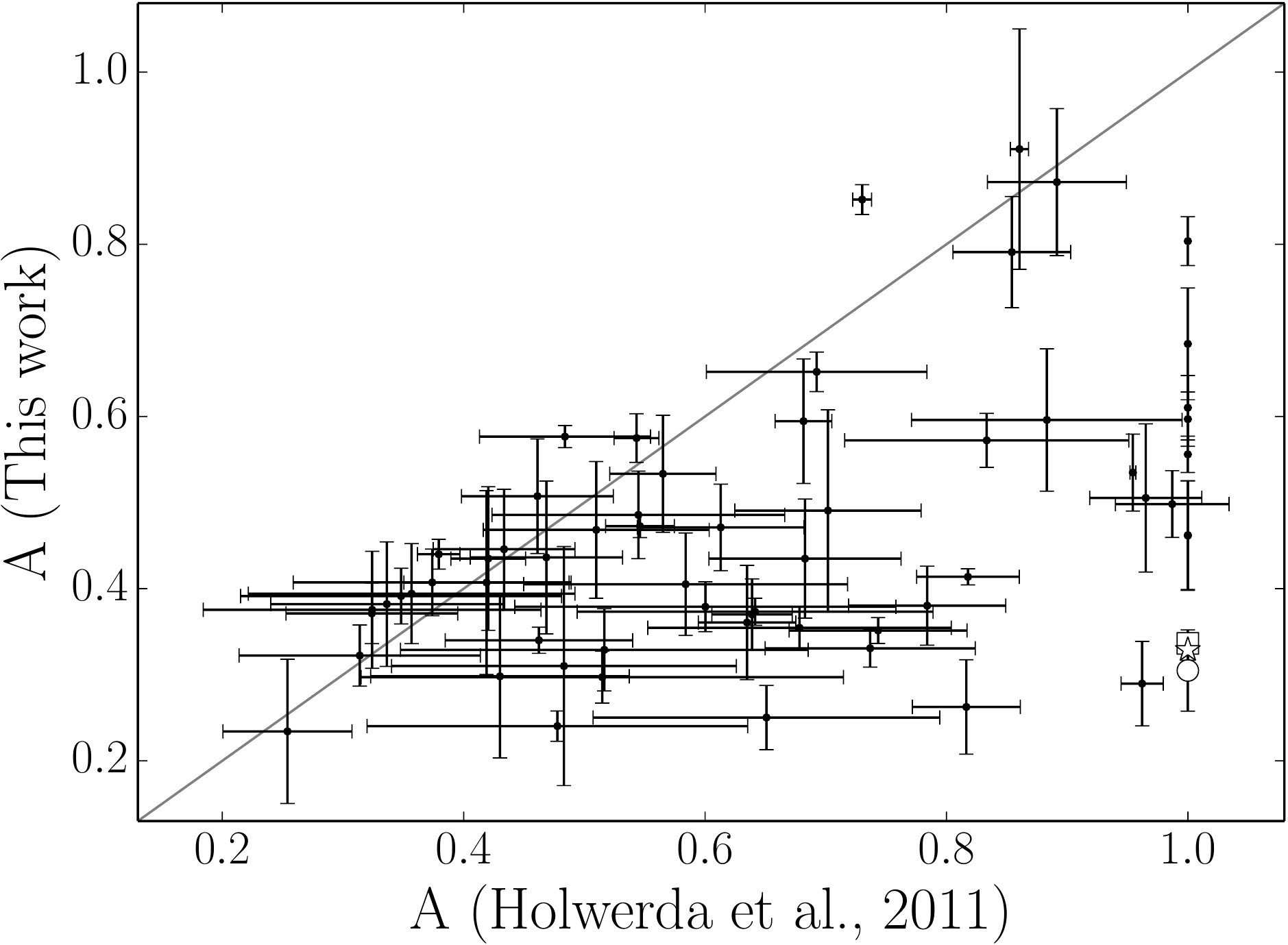}
  \end{center}
  \caption{Comparison of the values for the Asymmetry parameter for 141 WHISP galaxies. For the comparison we use a table with parameters provided with the online version of \protect\cite{Holwerda2011II}. Since the \textit{A} values in that table cover the range [0,2], we divided them by 2. \new{The vertical error bars have been estimated as the square root over the sum of the squared uncertainties resulting from the error in the central position of the galaxy and the errors in the flux values of the total \HI image. The horizontal error bars are calculated similarly with the values for the errors taken from the tables A1 and A2 of the online version of \protect\cite{Holwerda2011II}.}
  The values estimated in \protect\cite{Holwerda2011II} are higher or approximately equal to the values that we estimate. We show the moment zero images as well as the differences for the rotated images in Fig.\ \ref{asymmetry_comparison_Holwerda2011_examples} for the examples that have been marked with the square, star and open circle symbols. We could not determine a systematic effect which could have led to the significantly higher values in \protect\cite{Holwerda2011II}. }
  \label{asymmetry_comparison_Holwerda2011}
\end{figure}

Our values for the \textit{A} parameter already show a strong deviation from the results in \cite{Holwerda2011II} (see Fig.\ \ref{asymmetry_comparison_Holwerda2011}). There is a large number of galaxies for which we measure a lower \textit{A}. \new{The horizontal error bars are taken from the tables A1 and A2 in the online version from \cite{Holwerda2011II}. There is no further description of these errors with the online material. Thus, considering the error estimation in \cite{Holwerda2014}, we assume that the errors given in the online tables are the uncertainties stemming from the error in the central position as well as the flux values in the total \HI images. For the error bars in Fig.\ \ref{asymmetry_comparison_Holwerda2011} we took the square root of the sum of the squared errors. The vertical errors are generated in a similar way. The component that results from the uncertainty in the central position has been estimated following \cite{Holwerda2014} who vary the centre within a Gaussian with an FWHM of three pixels.} 

\begin{figure}
  \begin{center}
    \includegraphics[width=0.45\textwidth]{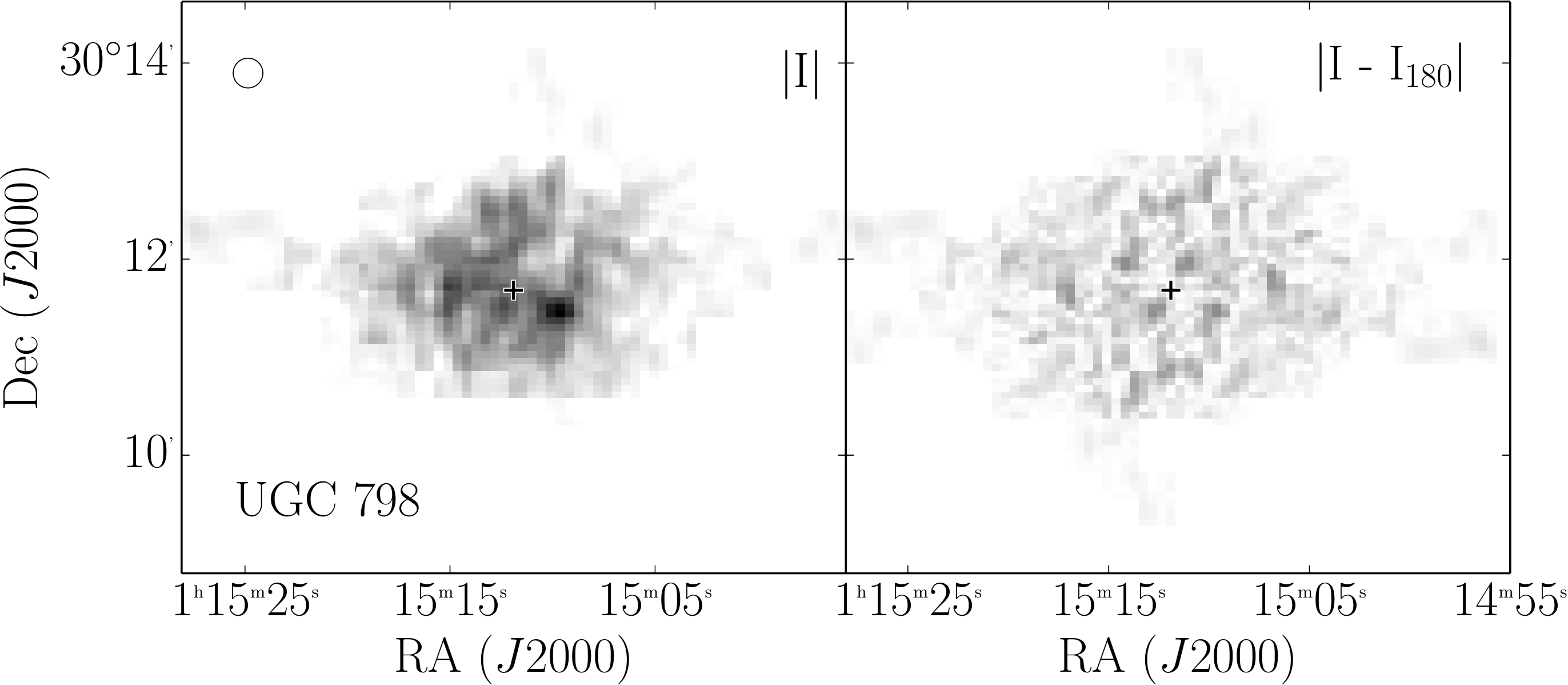}\vspace{0.2cm}
    \includegraphics[width=0.45\textwidth]{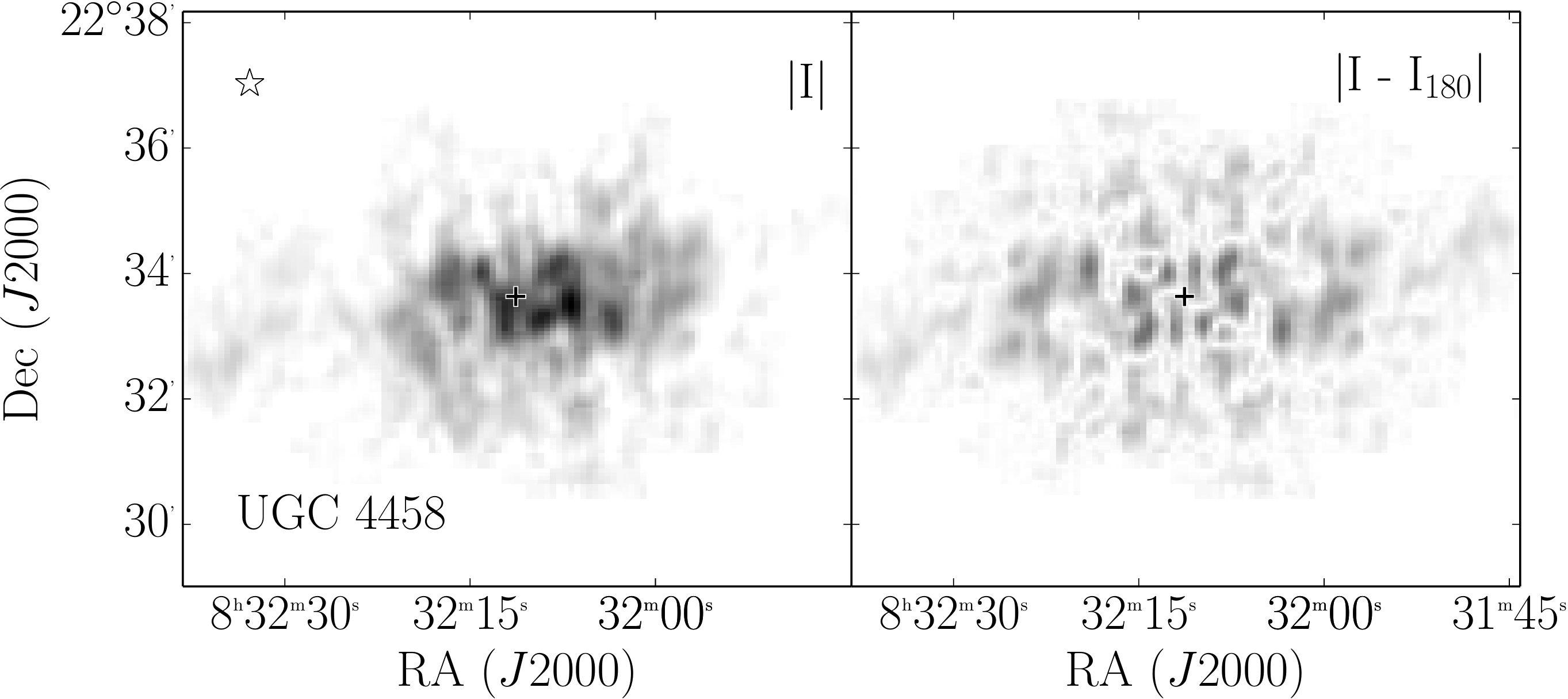}\vspace{0.2cm}
    \includegraphics[width=0.45\textwidth]{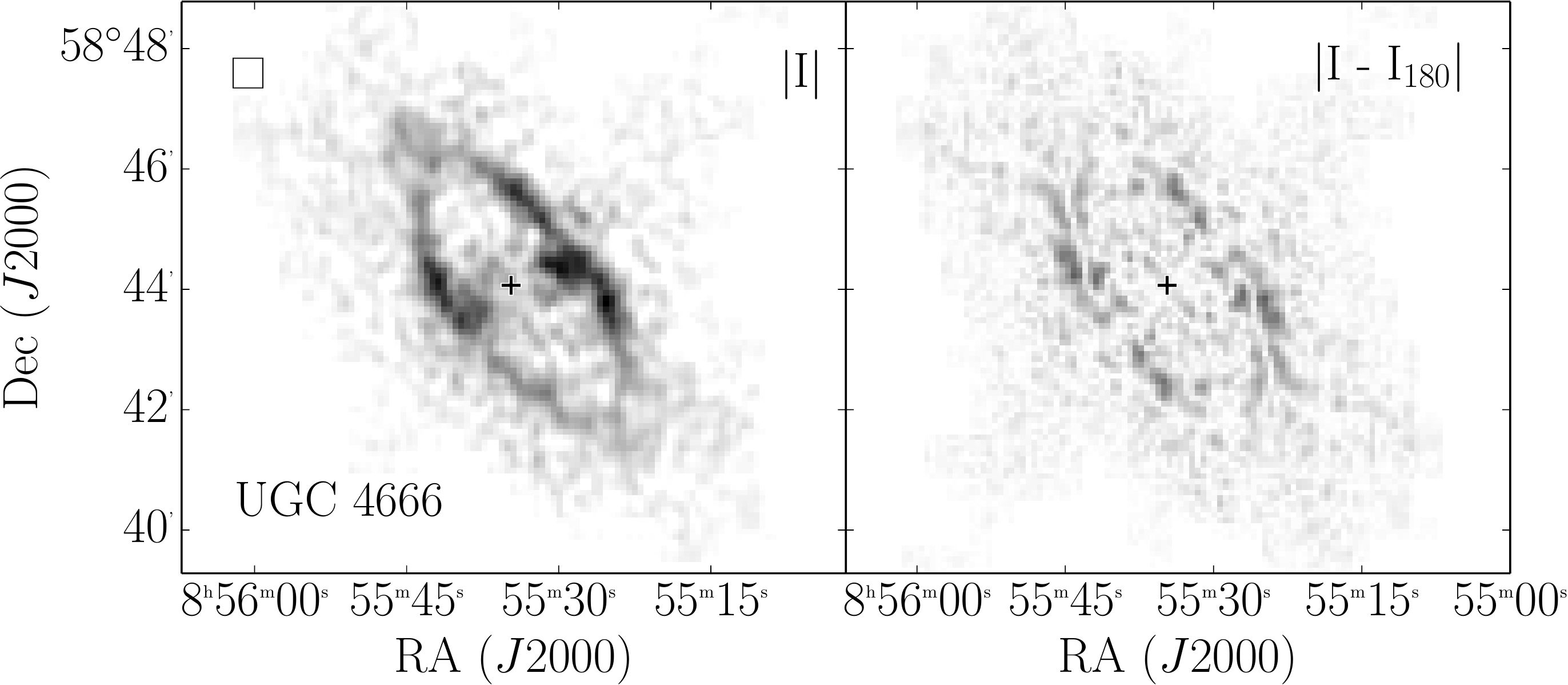}
  \end{center}
  \caption{Moment zero images (left panels) and difference images (right panel) for the moment zero image and its 180 degree rotated version.
  The black crosses indicate the optical centres of the galaxies. The Asymmetry values measured in \protect\cite{Holwerda2011II} for all three examples have a maximum of 1, which indicates that the galaxies are highly asymmetric. The difference do not show signs for strong asymmetries.}
  \label{asymmetry_comparison_Holwerda2011_examples}
\end{figure}

For a simple illustration of why we obtain lower values we selected three galaxies which have the highest possible \textit{A} parameter according to the analysis in \cite{Holwerda2011II} and for which we estimate a significantly lower \textit{A}. These galaxies are UGC 798, UGC 4458 and UGC 4666. Their \textit{A} values have been marked with special symbols in Fig.\ \ref{asymmetry_comparison_Holwerda2011}. In Fig.\ \ref{asymmetry_comparison_Holwerda2011_examples} we show the moment zero map (left panel) as well as the absolute difference between the moment zero map and its 180 degrees rotated version (right panel) for all three galaxies. The intensities in both panels are on the same scale for each of the three galaxies. Taking equation \ref{eqn_asymmetry} into account, the panels showing the difference must have an integrated intensity which is twice as high as for the moment zero map in order for the galaxy to have the maximum \textit{A} of 1. However, this is not the case for any of the examples. 

We suspect that the errors in \textit{A} are due to the use of the wrong optical centres for the objects, since this would explain not only the discrepancy in the Asymmetry but also other parameters like Concentration, M$_{20}$ and \newtwo{GM}, which will be defined below. In particular, the extreme values of $A=1$ could be explained with the centre of the galaxy being located outside of the mask defining the galaxy. We tried to reproduce the Asymmetry results from \cite{Holwerda2011II} by using the \HI flux weighted centre instead of the optical, swapping the pixel coordinates of RA and Dec, assuming the wrong Epoch while converting from RA and Dec to pixel coordinates, not excluding other objects that are in the same integrated \HI image and selecting a centre at random. However, none of the approaches reproduces the Asymmetry values consistently. 
\new{An explanation for the extreme \textit{A} values can be given for a few galaxies which have \HI images using mJy as a unit for the density rather than Westerbork Units (1 W.U. = 5 mJy). In these cases the thresholds are too high because the wrong unit is assumed leaving zero pixels for the evaluation of the Asymmetry parameter and hence giving \textit{A} values of $1$. UGC 798 and UGC 4666, which are shown in Fig.\ \ref{asymmetry_comparison_Holwerda2011_examples}, are among those galaxies.}

In addition to the Asymmetry parameter, \cite{Holwerda2011II} included the parameters Concentration, Smoothness, Gini, $M_{20}$, and \newtwo{the} newly defined parameter \newtwo{\textit{GM}}. The Concentration parameter is usually estimated as the logarithm of the ratio between the radii containing 80\% and 20\% of the total flux of an object \citep{Bershady2000}:
\begin{equation}
C = 5 \log{\frac{r_{80}}{r_{20}}}
\end{equation}
Since the \HI gas distribution in galaxies does in general have a lower central concentration than e.g. the stellar counterpart, the Concentration parameter estimated from total \HI images will have lower values than the ones estimated from optical images. \cite{Holwerda2012} \new{point out} that there had been an error in the Concentration estimation in a previous paper. This should be taken into account when evaluating the results presented in \cite{Holwerda2011II}.

The Smoothness parameter is defined as the difference between the original image and the smoothed version and has been developed by \cite{Conselice2003} as a clumpiness parameter to evaluate the degree of small scale structures. In this work we only estimate the Smoothness parameter for the WHISP galaxies in order to compare with \cite{Holwerda2011II}. However, since the choice of smoothing kernel necessary for its measurement is arbitrary, we exclude this parameter from further analysis.

The Gini parameter, which was first used on astronomical data by \cite{Abraham2003}, has its origin in economics, where it is used to describe the inequality of the distribution of wealth among a \new{nation's citizens}. Expressed in the context of images it is defined as the mean of the absolute difference between all combinations of pixels intensities $I_i$. Its limits are zero, if all pixels have the same intensity,
and one, if all intensity is contained in one pixel. The Gini parameter is defined as
\begin{equation}
\textrm{Gini} =  \frac{1}{\bar{I} n (n-1)}\sum_i{(2i-n-1)I_i},
\label{eqn_gini}
\end{equation}
where $\bar{I}$ is the mean pixel intensity and $n$ the number of pixels. The values $I_i$ have to be sorted in an ascending order. 

In addition to the Gini parameter, \cite{Lotz2004} introduce the moment of light parameter $M_{20}$. For every pixel, a second order moment is defined as the product of the pixel intensity and the squared distance from the centre of the galaxy:
\begin{equation}
M_i= I_i \left[ (x_i-x_c)^2 + (y_i-y_c)^2 \right].
\label{eqn_Mi}
\end{equation}
The $M_{20}$ parameter is calculated as the fractional sum over the second order moments of the brightest pixels containing 20\% of the total intensity and the sum over all second order moments:
\begin{equation}
M_{20}= \log_{10}{\frac{\sum_{i}{M_i}}{M_{tot}}}, \textrm{where} \sum_i{I_i}<0.2 I_{tot}.
\end{equation}
It traces bright structures, especially if they are offset from the centre. In optical images the $M_{20}$ parameter is therefore sensitive to structures like spiral arms, bars and phenomena associated with
merger processes, like e.g. multiple nuclei. Additionally, \cite{Holwerda2011II} introduce another parameter, which is a combination of the Gini and the $M_{20}$ parameters, called \new{ \textit{GM} }:
\begin{equation}
GM = \frac{1}{\bar{M} n (n-1)}\sum_i{(2i-n-1)M_i}.
\end{equation}
Comparing this definition to equations (\ref{eqn_gini}) and (\ref{eqn_Mi}),  \textit{GM}  is the Gini parameter for the distribution of second order moments $M_i$. 

\begin{figure}
  \begin{center}
    \includegraphics[width=0.45\textwidth]{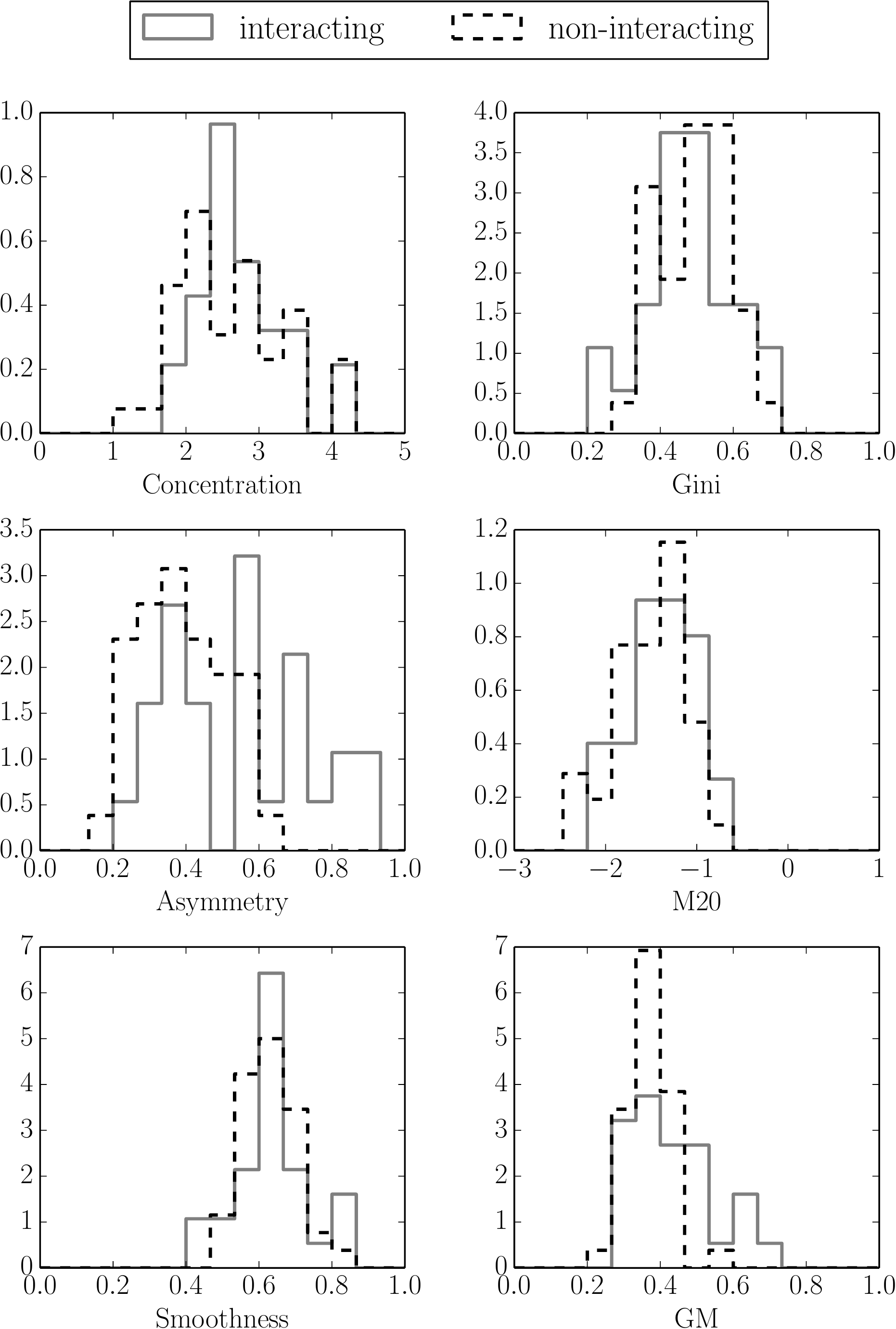}
  \end{center}
  \caption{Concentration, Asymmetry, Smoothness, Gini, $M_{20}$ and \textit{GM} for the visually inspected WHISP subsamples from \protect\cite{Noordermeer2005} and \protect\cite{Swaters2002} - \new{There is a small trend in the Asymmetry parameter with the distribution of interacting galaxies reaching higher values than non-interacting galaxies.}}
  \label{hist_int}
\end{figure}

We estimated all parameters for the WHISP data set \new{to see if the results from \cite{Holwerda2011II} can be confirmed. We used the visual classifications from \cite{Noordermeer2005} and \cite{Swaters2002} to divide the combined sample into interacting and non-interacting objects. \newtwo{It should be noted that galaxies that are classified as non-interacting do not show any sign of interactions. This, however, does not proof that they are not interacting. For convenience and to be consistent with the \cite{Holwerda2011II} notation, we will use the term non-interacting for galaxies that do not show any signs of interaction.}
Fig.\ \ref{hist_int} shows the distribution of morphometric parameter values for these two different groups. We find that a number of interacting galaxies can be roughly separated from the non-interacting sample using $A>0.65$ or $GM>0.6$. However, this separation is of a different nature compared with the findings in \cite{Holwerda2011II}. We find that the distribution of the Asymmetry parameter values for the interacting sample spans a range that exceeds the range of the non-interacting sample. The \new{ \textit{GM} } parameter shows a similar, but less pronounced feature. The results from \cite{Holwerda2011II} do not show these distinct parameter ranges. However, they find a much clearer separation between the peaks of the two distributions in the Asymmetry as well as \new{ \textit{GM} } parameter.}

\begin{figure*}
  \begin{center}
    \includegraphics[width=0.6\textwidth]{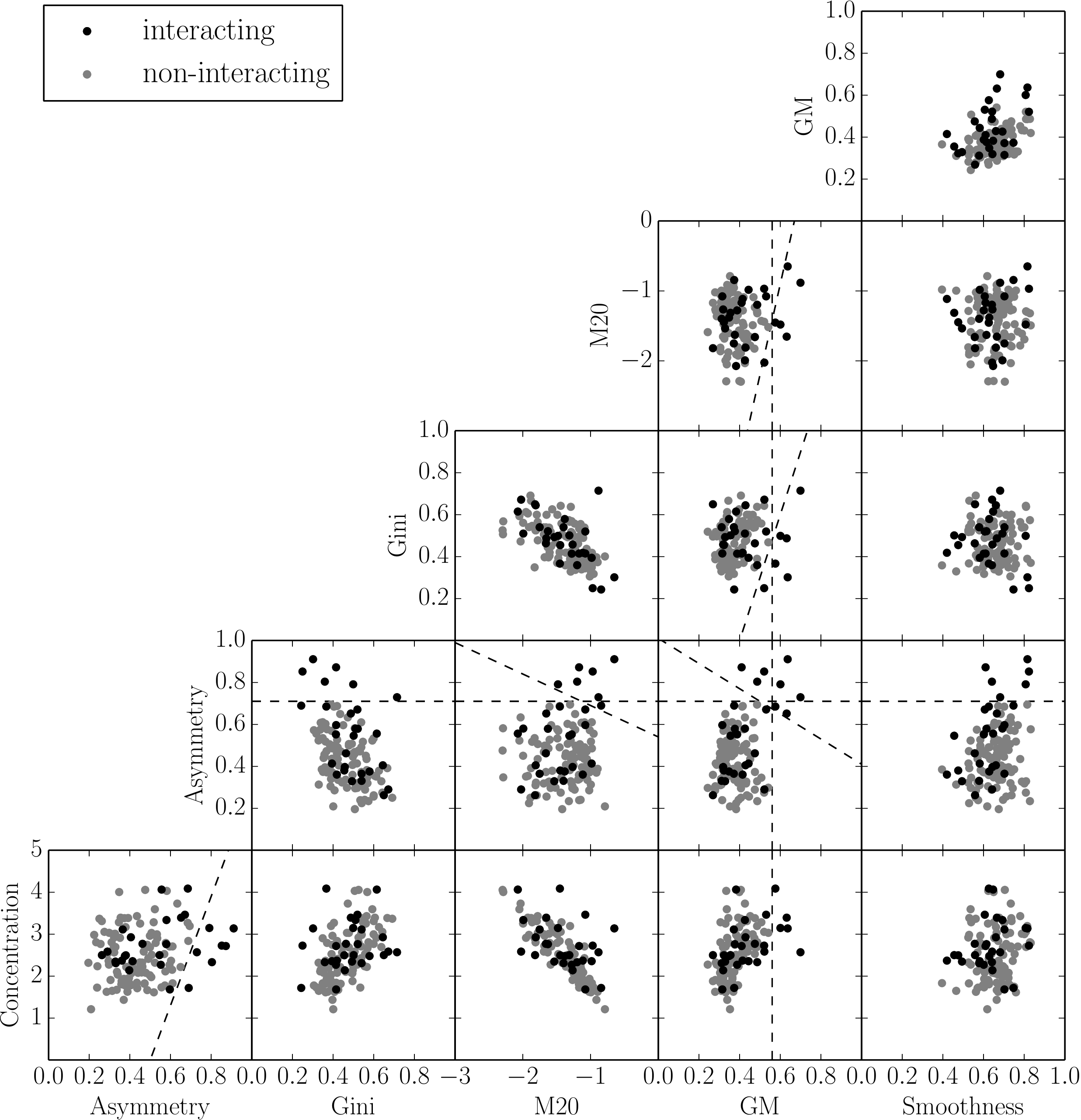}
  \end{center}
  \caption{Concentration, Asymmetry, Gini, $M_{20}$,  \textit{GM}  and Smoothness for the visually inspected WHISP subsamples from \protect\cite{Noordermeer2005} and \protect\cite{Swaters2002} - We do not find the same trends in parameter space as indicated in \protect\cite{Holwerda2011II}.}
  \label{corr_int}
\end{figure*}

\new{
Fig.\ \ref{corr_int} shows 2D projections of the distribution of the interacting and non-interacting subsamples in the space defined by all morphometric parameters (see \citealt{Scarlata2007} for the first example of such plot). In this space we can define a number of criteria to find interacting galaxies without including any non-interacting objects. The criteria from Table \ref{table:criteria} are displayed as dashed lines in Fig.\ \ref{corr_int}.
}

\begin{table}
  \caption{Number and fraction of interacting WHISP galaxies from the combined \protect\cite{Noordermeer2005} and \protect\cite{Swaters2002} set that have been correctly identified as interacting  using the respective criterion.}
  \begin{tabular}{p{0.1cm} p{3.2cm}|r}
    \multicolumn{2}{p{3.3cm}|}{Selection criterion}  & number (fraction) of galaxies  \\ \hline
    (a) & $GM>0.56$                    &  5 (18\%) \\
    (b) & $A>0.71$                     &  6 (21\%) \\
    (c) & $A>-0.15\times M_{20}+0.54$  &  7 (25\%) \\
    (d) & $M_{20}<13\times GM-8.7$     &  7 (25\%) \\
    (e) & $A>-0.6\times GM +1.01$      &  7 (25\%) \\
    (f) & $C<13\times A -6.5$          &  8 (29\%) \\
    (g) & (a)$\vee$(b)$\vee$(c)$\vee$(d) & 10 (36\%)\\
  \end{tabular}
  \label{table:criteria}
\end{table}
\new{
Although an individual classification returns better results for the \cite{Noordermeer2005} sample, we chose to use the combined \cite{Noordermeer2005} and \cite{Swaters2002} sample to cover a large range of galaxy properties and data qualities. A combination of the Asymmetry, \new{ \textit{GM} } and M$_{20}$ parameters is sufficient to recover $\sim$36\% of the galaxies that have been visually classified as interacting without including any of the non-interacting objects. A principle component analysis helps making the separation between the objects that can be classified as interacting in the original parameter space from the remaining galaxies more pronounced. The increase in the number of galaxies that can be correctly classified as interacting is, however, not significant. 
}

\begin{figure*}
 \begin{center}
  \includegraphics[width=0.4\textwidth]{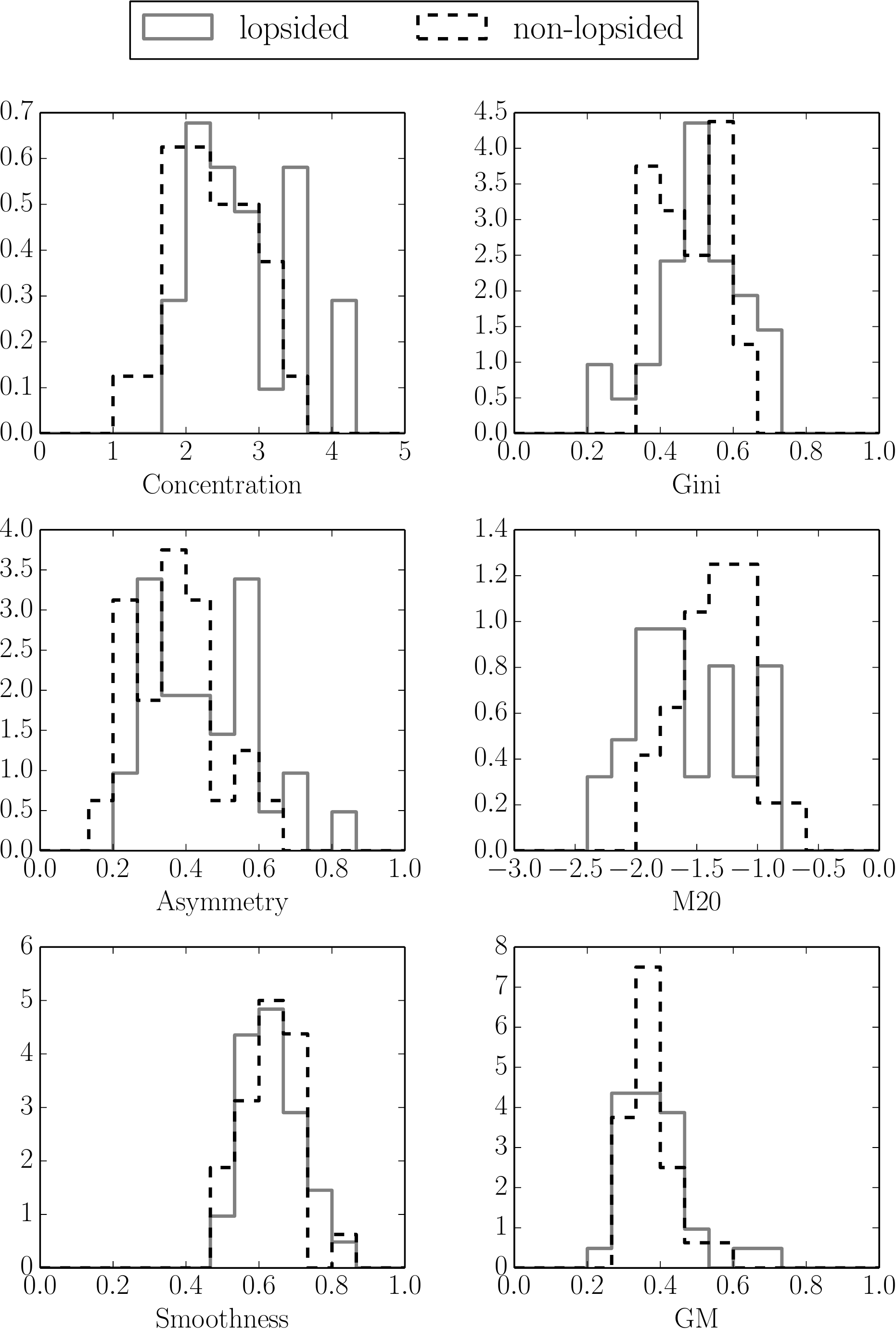}\hspace{0.1\textwidth}
  \includegraphics[width=0.4\textwidth]{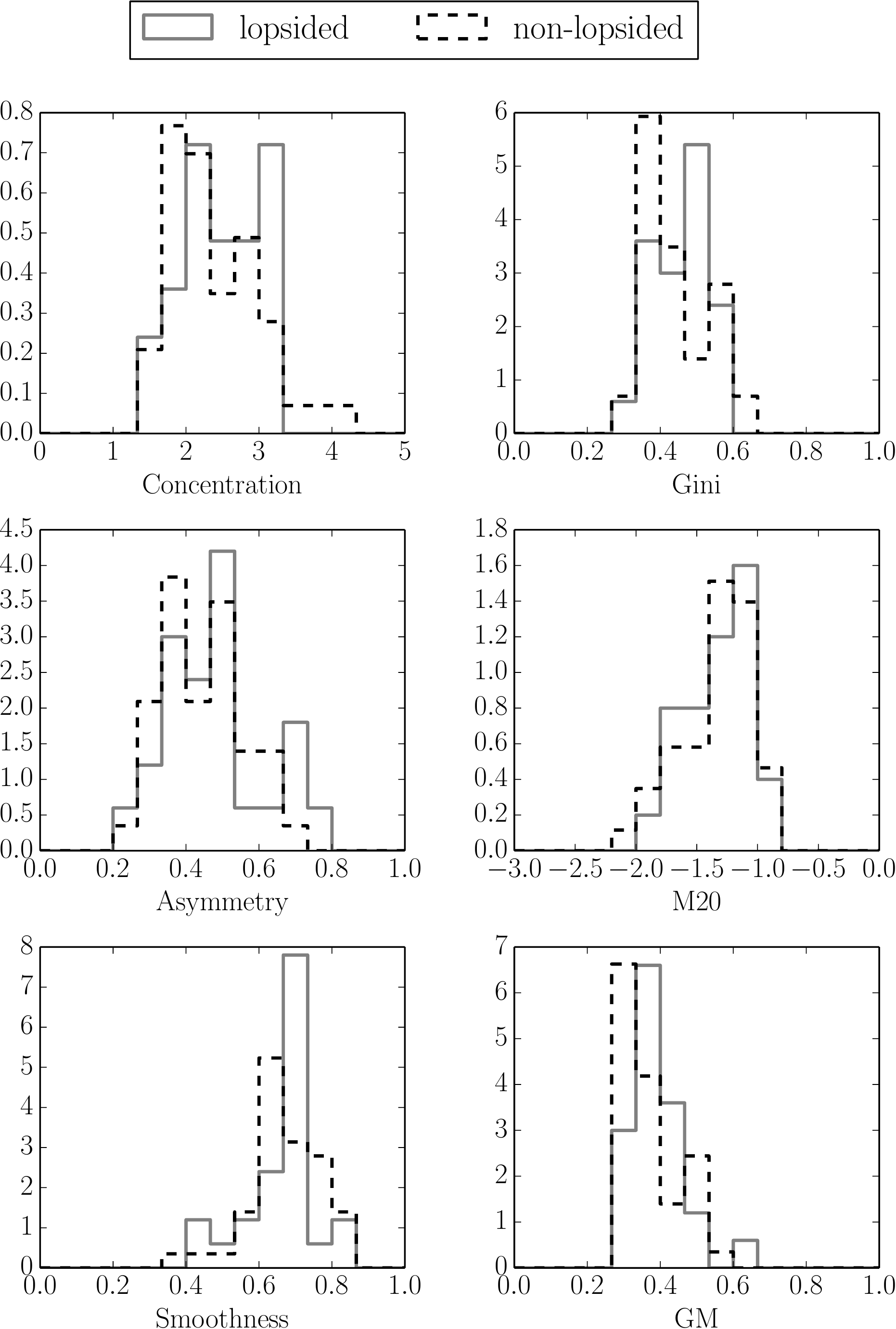}
 \end{center}
 \caption{Concentration, Asymmetry, Smoothness, Gini, $M_{20}$ and  \textit{GM}  for the visually inspected WHISP subsamples from \protect\cite{Noordermeer2005} (left panel) and \protect\cite{Swaters2002} (right panel).}
 \label{figure:histogram_lopsidedness}
\end{figure*}

\begin{figure*}
 \begin{center}
  \includegraphics[width=0.65\textwidth]{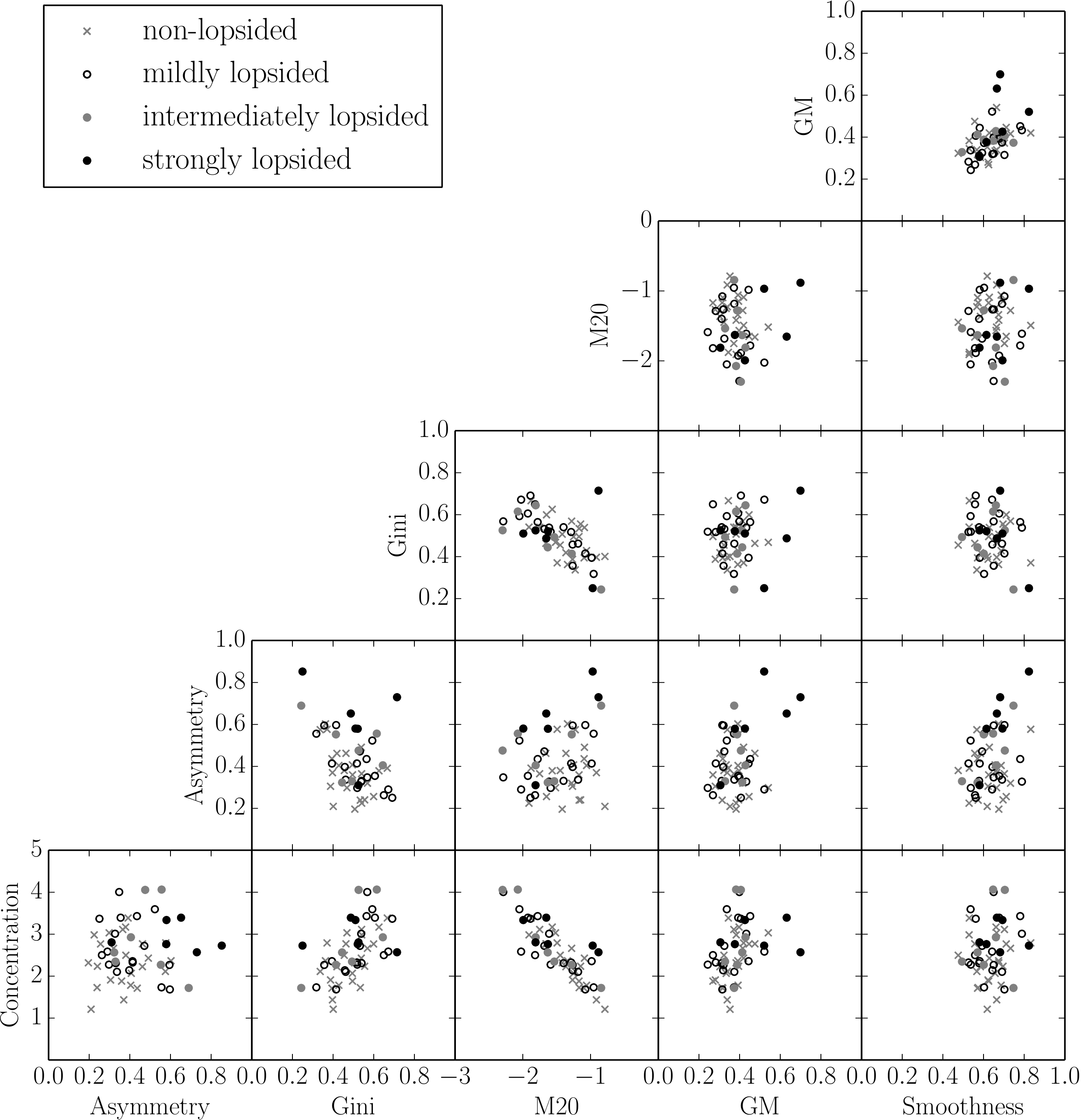}
 \end{center}
 \caption{Concentration, Asymmetry, Smoothness, Gini, $M_{20}$ and  \textit{GM}  for the visually inspected WHISP subsamples from \protect\cite{Noordermeer2005}.}
 \label{figure:corr_plot_lopsidedness_N}
\end{figure*}

\begin{figure*}
 \begin{center}
  \includegraphics[width=0.65\textwidth]{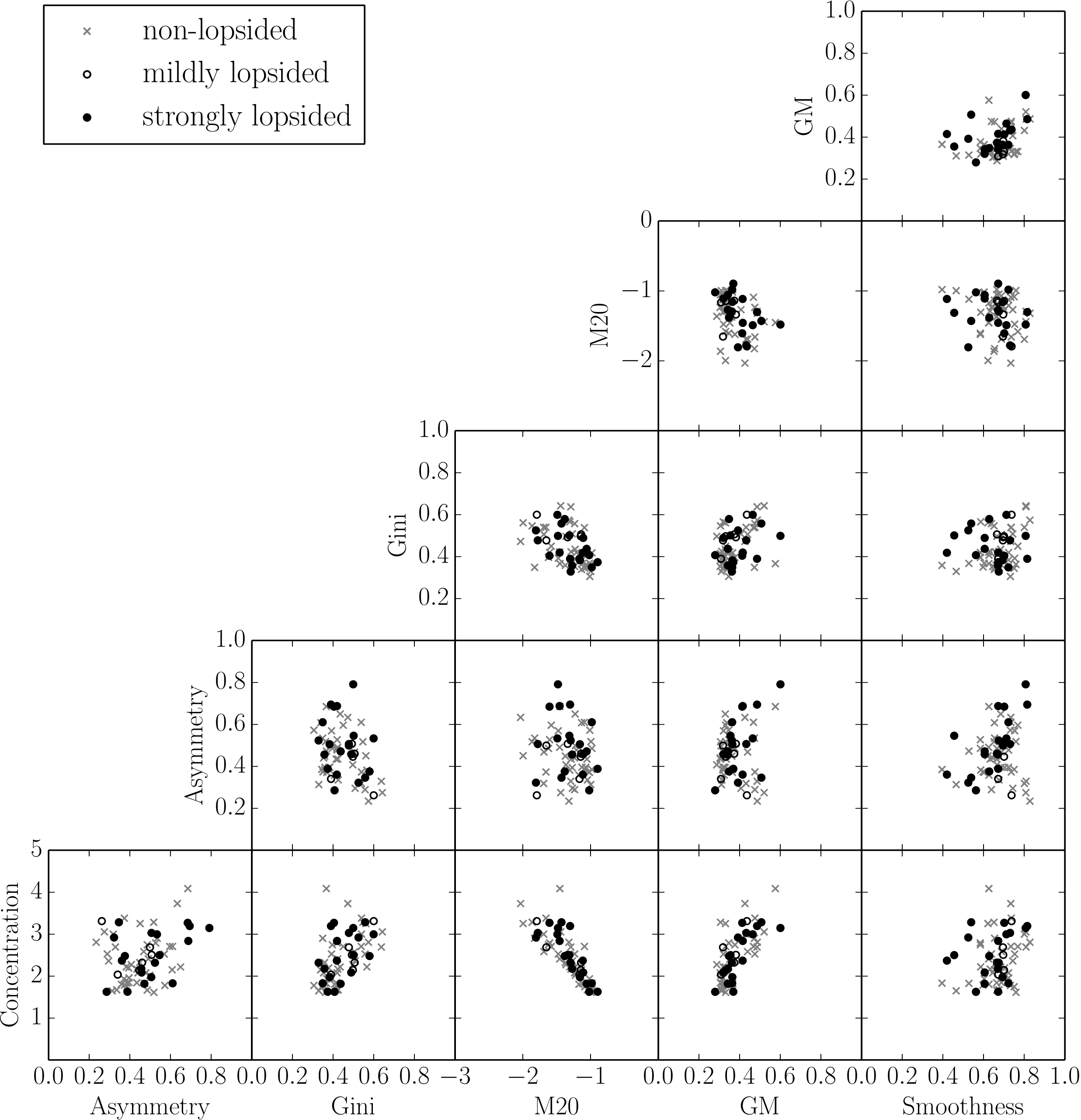}
 \end{center}
 \caption{Concentration, Asymmetry, Smoothness, Gini, $M_{20}$ and  \textit{GM}  for the visually inspected WHISP subsamples from \protect\cite{Swaters2002}.}
 \label{figure:corr_plot_lopsidedness_S}
\end{figure*}

\new{
Another interesting aspect for the use of these parameters is the classification of lopsidedness in galaxies
\newtwo{(please see \cite{Jog2009} for an overview of the lopsidedness phenomenon)}.
Since \cite{Noordermeer2005} and \cite{Swaters2002} use different scales for the strength of lopsidedness, we chose to analyse both sets separately. The results are shown in Fig.\ \ref{figure:histogram_lopsidedness}, \ref{figure:corr_plot_lopsidedness_N} and \ref{figure:corr_plot_lopsidedness_S}.}

\new{
There is no clear separation between the non-lopsided and the lopsided groups in the histograms. The multi parameter plots with the more detailed groups of lopsidedness (Fig.\ \ref{figure:corr_plot_lopsidedness_N} and \ref{figure:corr_plot_lopsidedness_S}) feature some strongly lopsided galaxies that are separated from the other groups in Asymmetry- \textit{GM}  space and in their higher  \textit{GM}  and Asymmetry values. The other groups of lopsidedness, however, show a very strong overlap among each other. Overall these trends are not sufficient to classify galaxies into different categories of lopsidedness. This might be attributed to the problem that different data properties like the signal to noise and the resolution strongly influence the values of the \new{morphometric} parameters, such as Asymmetry.
}

\begin{figure*}
 \begin{center}
  \includegraphics[width=0.7\textwidth]{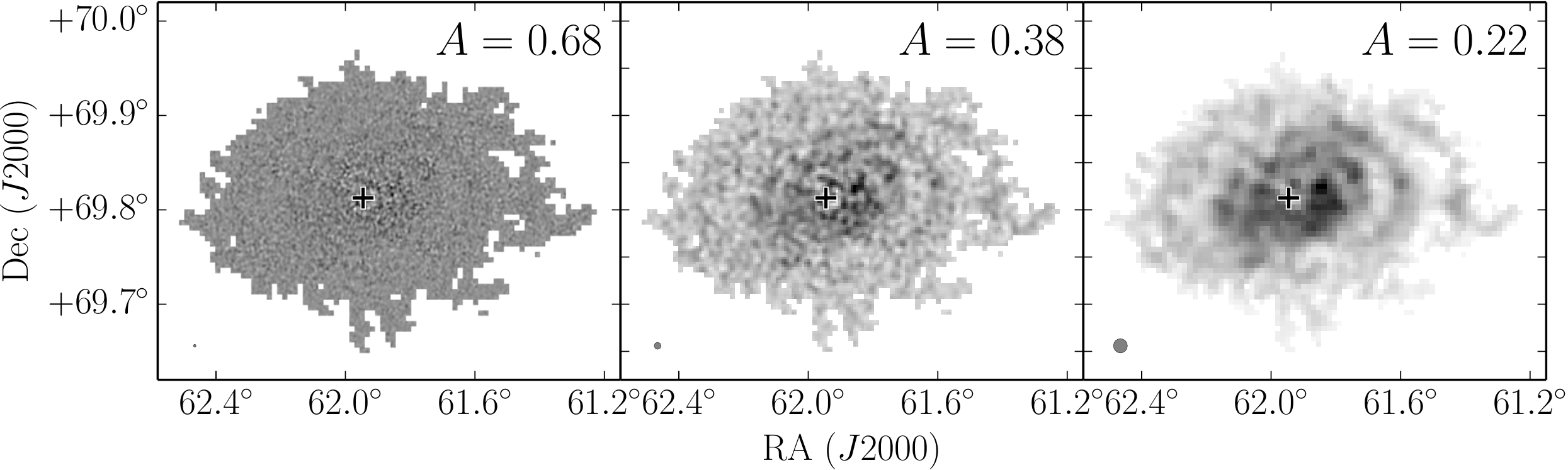}
 \end{center}
 \caption{The WHISP galaxy UGC 2953 at 3 resolutions (10.0''x11.2'', 
27.0'',54.2''). The black crosses indicate the optical galaxy center. The 
highest resolution image of this galaxy has a high asymmetry parameter although 
it is visually classified as intermediately asymmetric. With lower resolution the signal to noise increases and the asymmetry parameter decreases.}
 \label{ugc2953}
\end{figure*}

    An important problem in estimating \new{the Asymmetry parameter} becomes apparent when determining \new{it for} a galaxy at both different resolutions and different signal to noise ratios. The two effects are closely connected through the increase in signal to noise per pixel when smoothing and thereby lowering the resolution of an integrated \HI image. Using the three different resolution versions (12''/$\sin(\delta)$, 30'' and 60'') of galaxies from the WHISP survey, we noticed that with lower resolution the measured asymmetry decreases. An illustration of this effect is shown in Fig. \ref{ugc2953}. We picked UGC 2953 as an example of a low signal to noise case at highest resolution. We estimated the asymmetry using the optical centre as the reference point. The result for the asymmetry parameter changes from highly asymmetric ($A=0.68$) in the full resolution image to mildly/intermediately asymmetric ($A=0.22$) in the lowest resolution image. The latter value agrees better with the visual classification as non-lopsided by \cite{Noordermeer2005}. \new{This example also illustrates that the dynamic range in current observations of \HI disks is typically a few orders of magnitude less than in optical disks, affecting the sensitivity of the \textit{A} parameter.}

In the following sections we aim to show how certain image and galaxy properties influence the outcome of the measurement of the parameters mentioned above. 
To investigate the effects of signal to noise, resolution and inclination on the parameters used in \cite{Holwerda2011II} in more detail, we generated models using the tilted ring fitting code \tir \citep{Josza2007}, in which we could vary the mentioned properties independently.

\subsection{Model galaxies}

We used the tilted ring fitting code \tir to generate models of galaxies with varying intrinsic morphological asymmetry, inclination, and size relative to the angular resolution of the data.
Input parameters include circular velocity, surface brightness profile, position angle, inclination and scale hight for every simulated ring. We adopted the surface brightness profiles from \cite{Martinsson2011}

\begin{equation}
\Sigma_{\textrm{\tiny{\HI}}}(R) = \Sigma^{\textrm{\tiny{max}}}_{\textrm{\tiny{\HI}}} \cdot e^{-\frac{\left(R-R_{\Sigma,\textrm{\tiny{max}}}\right)^2}{2\sigma_{\Sigma}^2}},
\end{equation}
where $R_{\Sigma,\textrm{\tiny{max}}}=0.39 R_{\textrm{\tiny{\HI}}}$ denotes the position of the peak, $\sigma_{\Sigma}=0.35 R_{\textrm{\tiny{\HI}}}$ the width, $\Sigma^{\textrm{\tiny{max}}}_{\textrm{\tiny{\HI}}}$ the peak of the profile, and $R_{\textrm{\tiny{\HI}}}$ the \HI radius. 
We will use $R_{\textrm{\tiny{\HI}}}$ as the \HI scale length in units of modelled rings for the surface brightness profile and refer to it as the \HI radius hereafter. We chose the rotation curve representative for intermediate mass galaxies (see Fig.\ \ref{tirific_params}). 

\begin{figure}
 \begin{center}
  \includegraphics[width=0.36\textwidth]{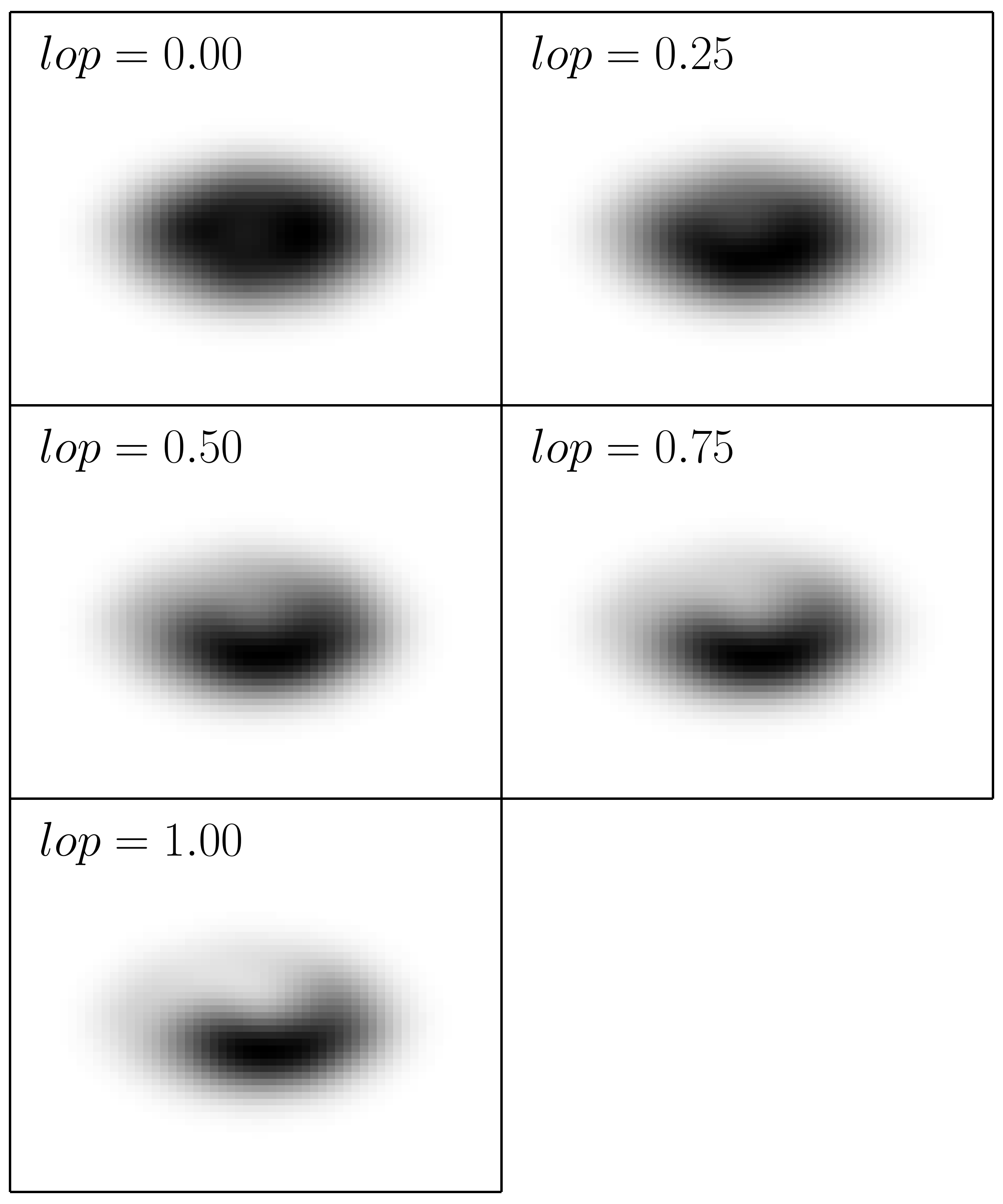}
 \end{center}
 \caption{Example of a \tir model galaxy with different levels of lopsidedness. An example set of input values for these galaxies are shown in Fig.\ \ref{tirific_params}.}
 \label{lop_example}
\end{figure}

To introduce morphological lopsidedness (asymmetries in the moment images), we added a harmonic surface-brightness distortion of first order to the models. These can be realized in \tir by choosing an amplitude and a phase for the distortion for every ring. An example of \tir input parameters can be found in Fig.\,\ref{tirific_params}. 
\new{The last two panels represent the values for the amplitude (SM1A) and the phase (SM1P) of the first order harmonic surface-brightness distortion. The values for the amplitude are chosen to be close to the surface brightness profile within the inner rings (except the innermost one). The strength of the added lopsidedness is then defined as a multiplier to the SM1A profile with a value of 0 representing symmetric objects and a value of 1 for strongly lopsided models. In the latter case all flux in the rings that are affected is azimuthally shifted using the phase of the harmonic surface-surface brightness distortion. An example for the different lopsidedness levels is shown in Fig.\ \ref{lop_example}. Using this definition, the strongest lopsided galaxy does not have a maximum Asymmetry value of 1. This value is, however, usually not reached for real galaxies, where the Asymmetry values usually span a range between 0 and 0.6.}

All our initial models consist of 11 rings. For the variation of the resolution of our models, we generated high resolution models, where the beam size corresponded to the sampling size. To achieve a certain angular galaxy size and hence resolution in beams across the major axis we first re-gridded these models to the desired angular size and then ``observed'' them by applying the instrumental broadening function. Using this approach we assumed that our highest resolution models corresponded to the clean components of a real observation. The resulting cubes were passed to the source finding application \SoFiA \citep{Serra2015} to determine the mask of every galaxy, i.e. all pixels that contain signal.
\new{The method in \SoFiA used to define pixels that contain signal is the smooth and clip method. The 3D data cube is smoothed in all 3 dimensions at different scales and all pixels that lie above a 4$\sigma$ noise threshold at that particular resolution are added to the mask of the cube. The mask is a separate 3D cube of the same size as the cube containing the galaxy, with all pixels belonging to the galaxy having the value 1 and all other pixels being 0. The smooth and clip method is an established way of defining the extent of the \HI emission in galaxies and searches for emission at different scales, because it assumes that the signal to noise is highest when the smoothing kernel matches the scale of the emission. For the experiments done in the next few subsections masks} were generated from the noiseless cubes and later used on all versions with varying noise level. Here, we assumed, that with low signal to noise faint emission that is not detected within the actual source finding step can be recovered by growing the masks in a subsequent step. \new{This is of course an idealised approach, which fails if the signal to noise in the 3D cube becomes too low. We chose to take this approach nonetheless to be able to investigate the effect of increasing noise level without sections of the galaxy being cut off. Additionally, we did not choose a signal to noise threshold to restrict the pixels that entered into the calculation.} 
We added noise using a noise cube that was generated from the Stokes Q component of an observation with the WSRT \citep[see][]{Serra2012II}. The noise cube has a resolution of 30''. Consequently, all our \tir models are smoothed to a round beam of 30''.

\subsection{Effect of S/N variations}\label{subsubsec:s2n}

\begin{figure}
 \begin{center}
  \includegraphics[width=0.36\textwidth]{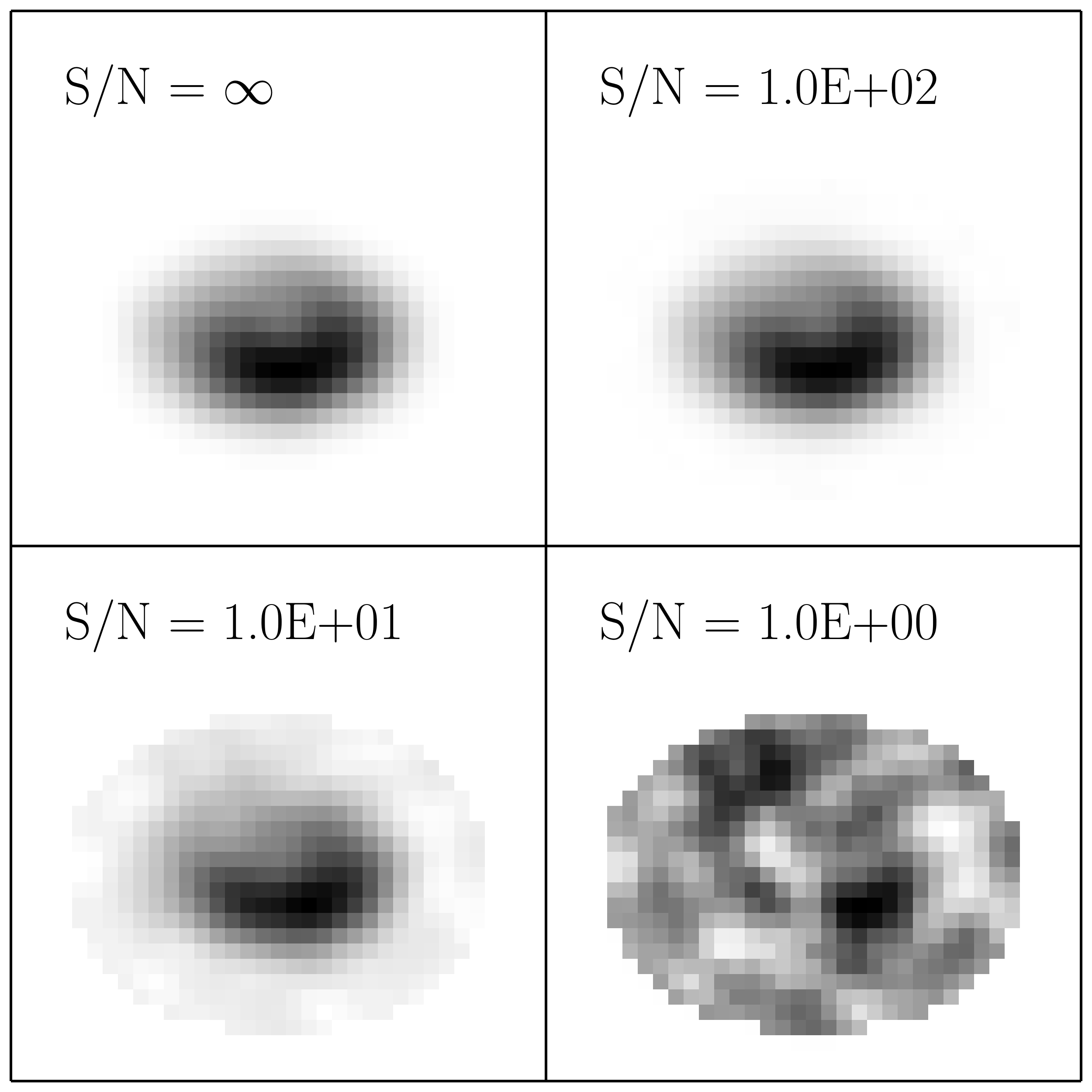}
 \end{center}
 \caption{Example of a \tir model galaxy with decreasing signal to noise level (S/N according to Eq.\ \ref{eq_s2n1} and \ref{eq_s2n2}). The input values for this galaxies are shown in Fig.\ \ref{tirific_params}. With decreasing S/N level the apparent lopsidedness is more and more replaced by asymmetric structures in the noise.}
 \label{s2n_example}
\end{figure}

To investigate the influence of signal to noise on the recovered parameters \textit{C}, \textit{A}, G, M$_{20}$, and \textit{GM} we generated noiseless model cubes for five galaxies with the same radial surface brightness density profile (see Fig. \ref{tirific_params}) and different intrinsic lopsidedness due to increasing amplitudes for the harmonic surface brightness distortion (see Fig.\ \ref{lop_example}). 
\new{We chose a resolution of 5 beams across the major axis. This choice is based on the results from \cite{Duffy2012}. Using simulations, they found that only 0.5\% of the galaxies expected in a shallow survey with Westerbork/APERTIF would have a resolution of $>5$ beams across the major axis. }
Fig.\ \ref{s2n_example} shows an example of one such galaxy which was generated using the parameters shown in Fig. \ref{tirific_params} and with different levels of noise added. For decreasing signal to noise the perceived lopsidedness is more and more determined by the noise instead of the intrinsic lopsidedness.

\begin{figure*}
 \begin{center}
  \includegraphics[width=0.9\textwidth]{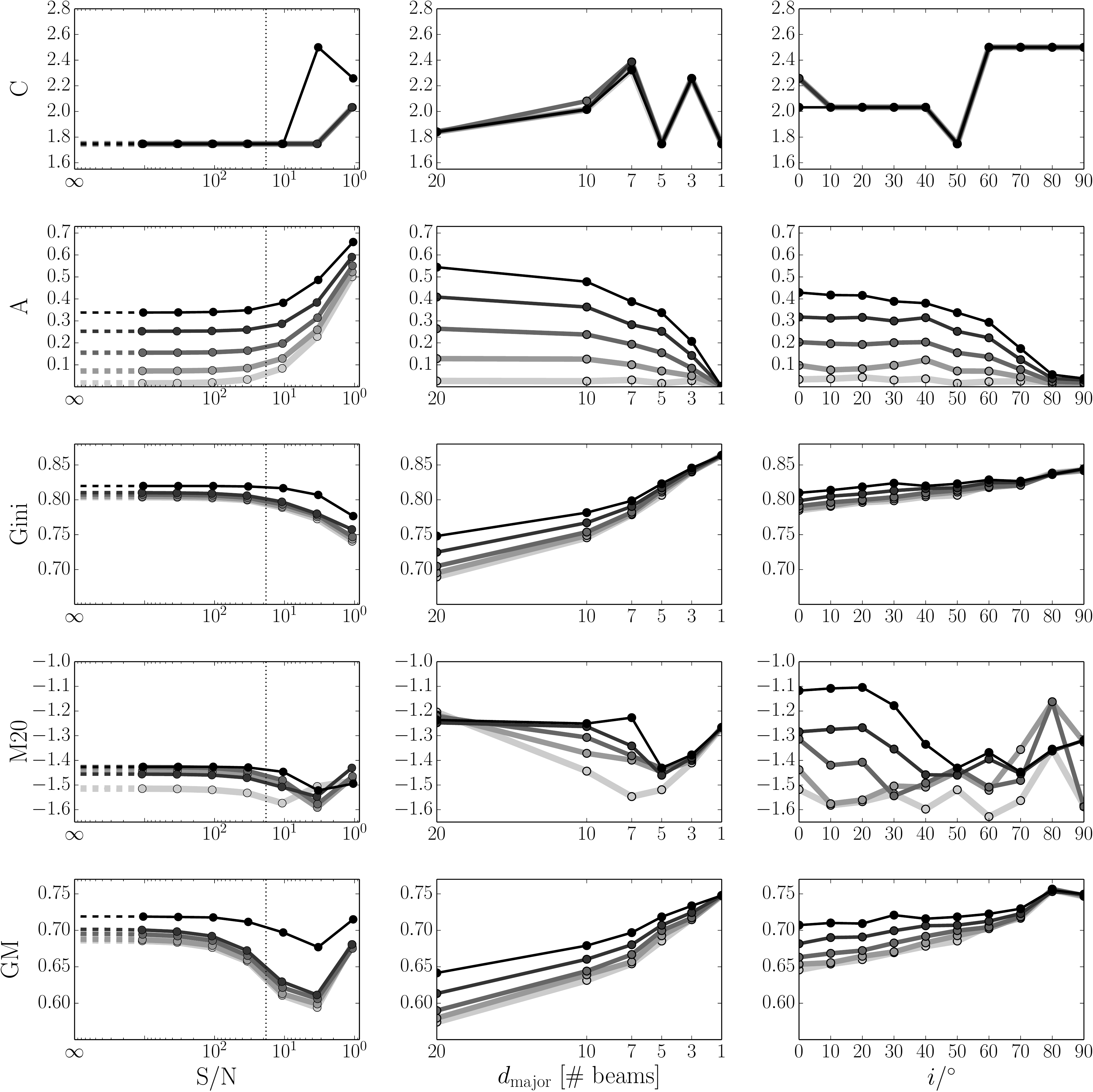}
 \end{center}
 \caption{Signal-to-noise, resolution and inclination dependence of \textit{C}, \textit{A}, Gini, $M_{20}$ and \textit{GM} for galaxy models with different intrinsic morphological lopsidedness. The grey level is indicating the strength of the lopsidedness with light grey for non-lopsided to black for highly lopsided models. The five different levels correspond to the 5 examples shown in Fig.\ \ref{lop_example}. All models were based on 11 rings. For every panel only one galaxy parameter was varied respectively ($S/N$,$d_\textrm{major}$ and $i$). The non-varying parameters chosen for the models were $S/N=\infty$, $i=50^{\circ}$, $d_\textrm{major} = 5$ beams.
 The dotted line indicates the maximum signal to noise for galaxies in the WHISP samle (Eq. \ref{eq_s2n2}). The most suitable parameter for the estimation of the degree of lopsidedness in a galaxy is the \textit{A} parameter. It shows a very good separation between the different lopsidedness models and features the slowest rate of change.}
 \label{dep_parameters}
\end{figure*}

Apart from the varying noise levels and the amplitude of the harmonic surface brightness distortion we kept all other galaxy properties, in particular the resolution and inclination, constant (see caption in Fig.\ \ref{dep_parameters}). The results for the signal to noise dependence of the \new{morphometric} parameters can be found in the first column of Fig.\ \ref{dep_parameters}. The dotted line indicates the maximum signal to noise in the WHISP data set. \new{Galaxies observed with Westerbork/APERTIF or ASKAP will have similar S/N characteristics.} The different shades of grey represent the degree of lopsidedness - with light grey for non-lopsided and black for highly lopsided galaxies.

It becomes immediately apparent that the Asymmetry parameter (second row) is the most promising parameter with regards to the separation of objects with different degrees of lopsidedness. For S/N values above 100 the recovered \textit{A} does not deviate significantly from the intrinsic value. All examples are well separated with more lopsided galaxies having higher \textit{A} values. Towards lower signal to noise values the measured \textit{A} increases. At a \new{S/N of about 3} the two galaxies with the lowest lopsidedness show similar \textit{A} values while the three intermediate to high lopsidedness cases can still be separated very well. For even lower S/N the the separation between all models becomes very small. The discovered trends show that it is not possible to compare \textit{A} values for galaxies with different S/N below 100. It is however noteworthy, that the Asymmetry parameter always deviates towards higher values. Furthermore, the slow and smooth increase for lower S/N values suggests that a correction of this noise bias is possible. We will present an approach for the correction in section \ref{section:bias}.

For S/N values above 100 the other parameters show very similar trends. Except for small variations they remain at the initial values for infinite signal to noise. The Concentration parameter does not show any significant separation between galaxies with different degrees of lopsidedness. This can be attributed to the way lopsidedness was introduced. The harmonic surface brightness distortion works on individual rings, thus the total intensity in annuli is conserved and the Concentration parameter remains unchanged regardless of the amplitude of the distortion. The Concentration parameter therefore does not seem suited to separate galaxies with different degrees of lopsidedness.

The Gini parameter shows at best a marginal separation between the four strongest lopsided models for S/N values above 100. For values below 100 the Gini value decreases as expected because with higher noise level the noise is dominating the flux values in each pixel and the distribution of the flux values is not restricted to the pixels containing the strongest signal of the galaxy.

The $M_{20}$ parameter does not show any meaningful separation of the 5 different lopsidedness models. It is not possible to distinguish between the intermediate and high asymmetry models. Only the lowest asymmetry example is well separated at lower $M_{20}$ values. For S/N values below 1 this trend however also disappears.

For high S/N values the \new{\textit{GM}} parameter exhibits a trend, where higher values correspond to higher lopsidedness in the model. However, already at S/N values above 100, the measured \new{\textit{GM}} parameter slightly decreases with decreasing S/N. For S/N values of the order of the WHISP galaxies models start to overlap in covered \new{\textit{GM}} ranges.

Based on Fig. \ref{dep_parameters} \textit{A} is the most useful parameter for the separation of galaxies with different levels of lopsidedness. It is superior due to the smooth trends with signal to noise and should be preferred provided that it is possible to correct the values for the positive offset, which we will refer to as the noise bias.

\subsection{Effect of resolution}\label{eff:resolution}

\begin{figure}
 \begin{center}
  \includegraphics[width=0.36\textwidth]{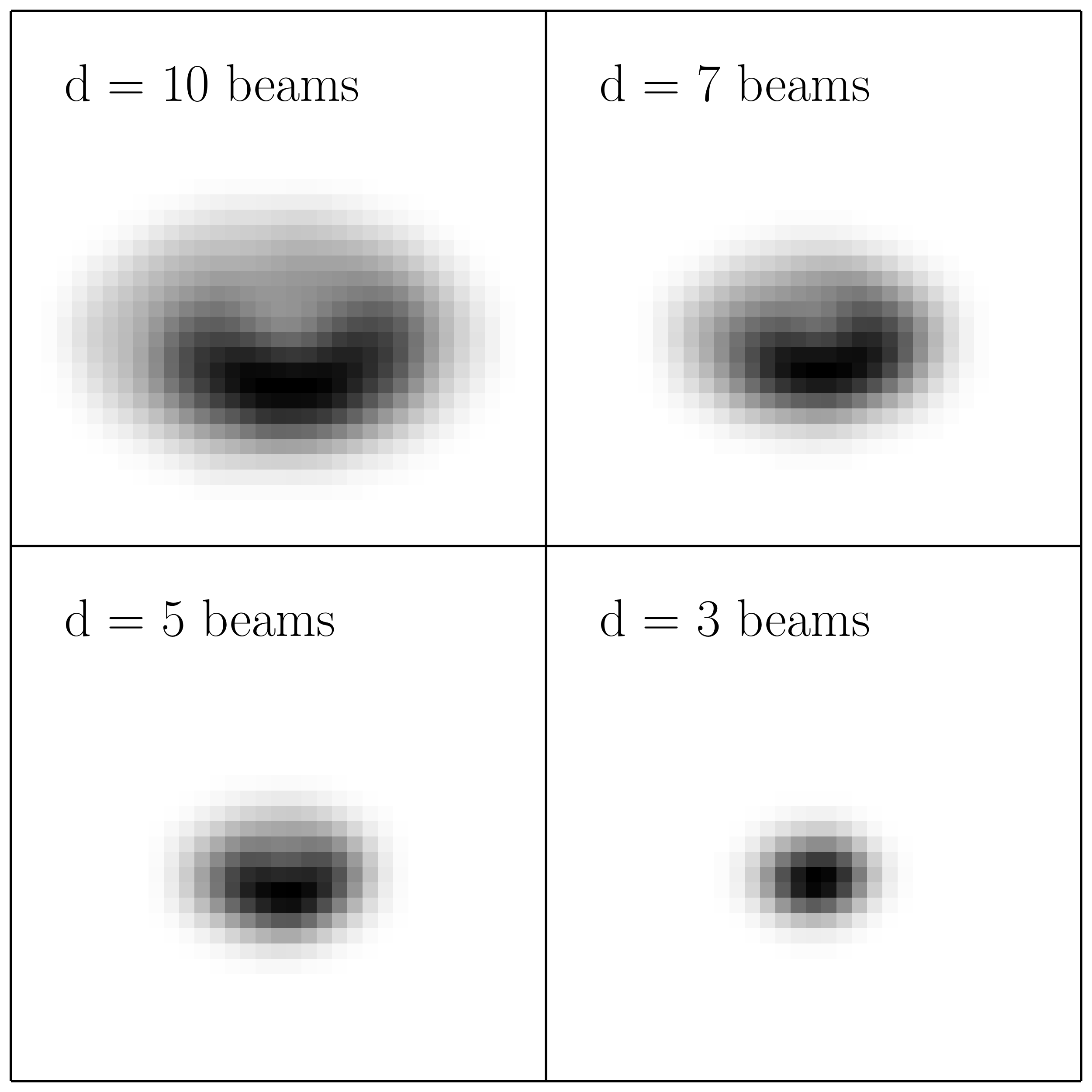}
 \end{center}
 \caption{Example of a \tir model galaxy with decreasing angular resolution. The input values for this galaxies are shown in Fig.\ \ref{tirific_params}. With decreasing angular resolution the apparent lopsidedness decreases with respect to the intrinsic one.}
 \label{res_example}
\end{figure}

We used the \tir models described above to investigate the influence of angular resolution on the estimation of the \new{morphometric} parameters. In this step we did not add noise to the models in order to allow us to separate the two effects. Fig.\ \ref{res_example} shows an example that has been set to resolutions from 20 to 3 beams across the major axis. We degrade the resolution of our models in terms of beam sizes across the major axis by re-gridding them on different pixel scales before convolving with a Gaussian beam.

Decreasing the resolution of the galaxies is the same as placing the models at \new{larger distances}. In Fig.\,\ref{dep_parameters} (second column) we show the results for the same galaxy examples as used in subsection \ref{subsubsec:s2n}. As before, the darker shades of grey correspond to higher intrinsic lopsidedness. 

The asymmetry parameter again turns out to separate the \new{different lopsidedness models} best. For lower resolution and thus examples at \new{larger distance} the measured \textit{A} value decreases compared to the intrinsic one. Down to a resolution of about three beams across the major axis all models are still separable, although models with lower degrees of lopsidedness start covering similar value ranges. Since a decrease of resolution is connected to a loss of information, it is not possible to correct the estimated \textit{A} for this effect. It is thus advisable to group galaxies into different classes of angular resolution for a comparison of the degree of lopsidedness among different objects. Another possibility is the decrease of the angular resolution of objects in a group to the value of the object with the lowest resolution. This would at the same time increase the S/N, which in turn decreases the influence of the noise bias on the \textit{A} parameter. For the large blind surveys that are planned for WSRT/APERTIF or ASKAP this approach is however not the best, considering that the majority of objects will have very low resolution ($\le 3$ beams across the major axis).

The Concentration parameter does neither show any significant separation between the different objects nor any trend with resolution in the measured value. Gini and \new{\textit{GM}} do show some separation between the models with the highest degree of lopsidedness for the highest resolutions (20 and 10 beams across the major axis). Towards lower resolutions this separation however decreases. The models with intermediate to low degrees of lopsidedness do not show any significant separation. The $M_{20}$ parameter does not consistently separate objects with different degrees of lopsidedness over the range of resolutions investigated.

\subsection{Effect of inclination}\label{eff:inclination}

\begin{figure}
 \begin{center}
  \includegraphics[width=0.36\textwidth]{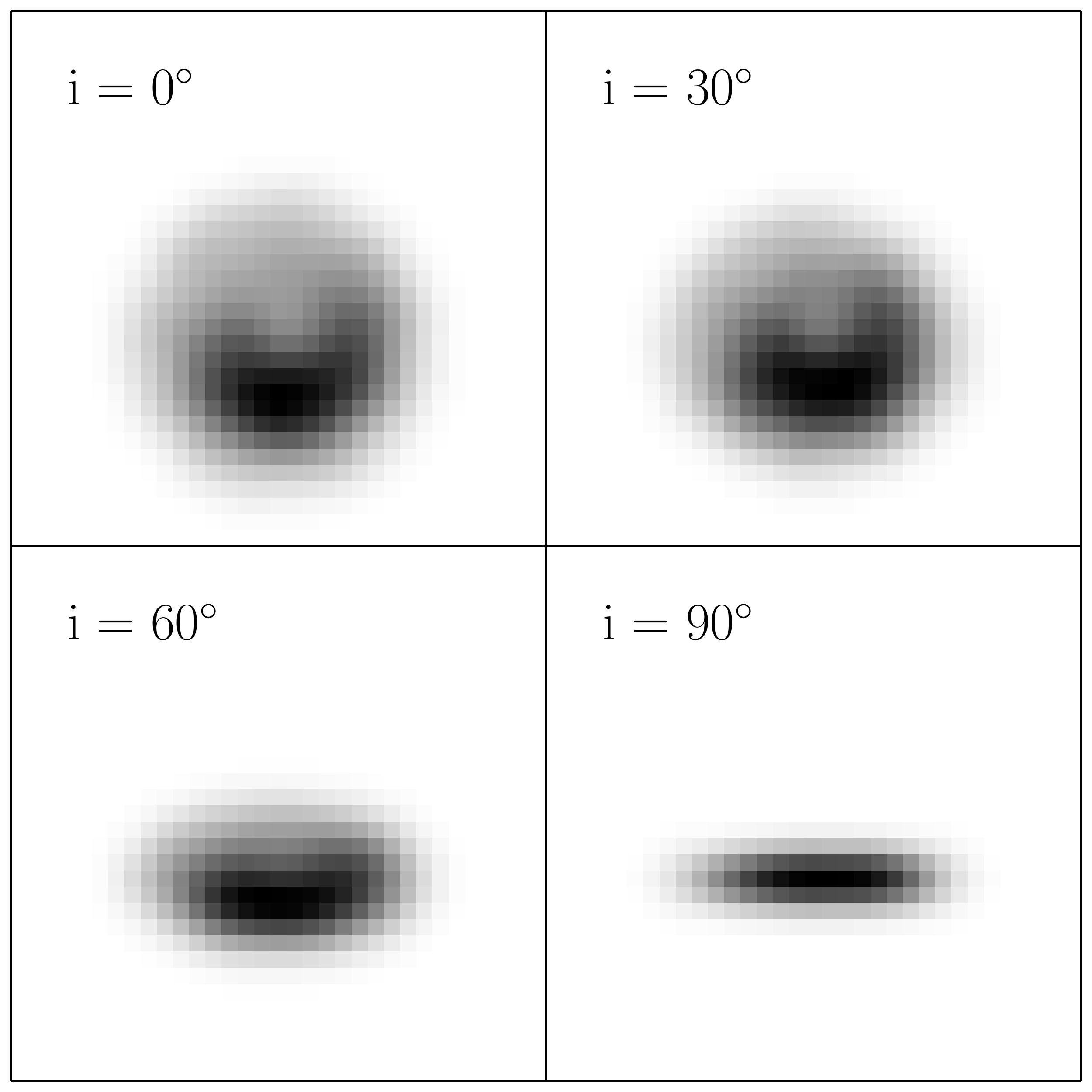}
 \end{center}
 \caption{Example of a \tir model galaxy with decreasing angular resolution. The input values for this galaxies are shown in Fig.\ \ref{tirific_params}. With decreasing angular resolution the apparent lopsidedness decreases with respect to the intrinsic one.}
 \label{incl_example}
\end{figure}

For the investigation of the effect of inclination on the \new{morphometric} parameters we used the same model approach as described in Sect.\ \ref{subsubsec:s2n} and \ref{eff:resolution}. For this experiment we fixed the resolution at five beams across the major axis and chose an infinite signal to noise.
All other galaxy properties are the same as in the previous sections. The only parameter that was varied was the inclination. Fig.\ \ref{incl_example} shows four galaxies at different inclinations varying from 0$^\circ$ to 90$^\circ$. With increasing inclination the apparent lopsidedness of the models appears to decrease.

For a detailed analysis we again used the same models with five different degrees of lopsidedness. The results are shown in the third column of Fig.\ \ref{dep_parameters}. When comparing the last two columns in Fig.\ \ref{dep_parameters} it seems that inclination and resolution have similar effects on the Asymmetry, the Gini and the \new{\textit{GM}} parameters except for a smoother trend for the resolution. This is not surprising as the change towards higher inclination is mainly a decrease of resolution along the minor axis of a galaxy. The parameter separating the different lopsidedness models best is again the Asymmetry parameter. All other parameters have similar limitations to the ones found in Sect.\ \ref{eff:resolution}.

Our findings for the Asymmetry parameter differ from results from previous work. \cite{Holwerda2011I} find that below an inclination of $80^{\circ}$ the measured \textit{A} does not suffer from inclination effects. In contrast, using \tir models we find that (1) the inclination limit for reliable \textit{A} measurement is lower and (2) the effect of inclination on the measured \textit{A} strongly depends on the azimuthal position of the lopsidedness within the galaxy.

\cite{Holwerda2011I} use existing integrated \HI images from the THINGS survey to investigate the dependence of \textit{A} on inclination. They rotated the images around the major axis to simulate galaxies with higher inclination and rebinned the images to the initial pixel size. Thereby they assumed that the galaxies featured an infinitely thin disk. This assumption however does not hold for the majority of galaxies. To demonstrate the difference between this 2D and a full 3D approach, we generated \tir model galaxies with increasing inclination and degree of lopsidedness. The azimuthal position of the lopsidedness was chosen at an intermediate position between major and minor axis. For the sake of comparability with \cite{Holwerda2011I}, we chose an angular resolution of 20 beams across the major axis for our models.

\begin{figure}
 \begin{center}
  \includegraphics[width=0.45\textwidth]{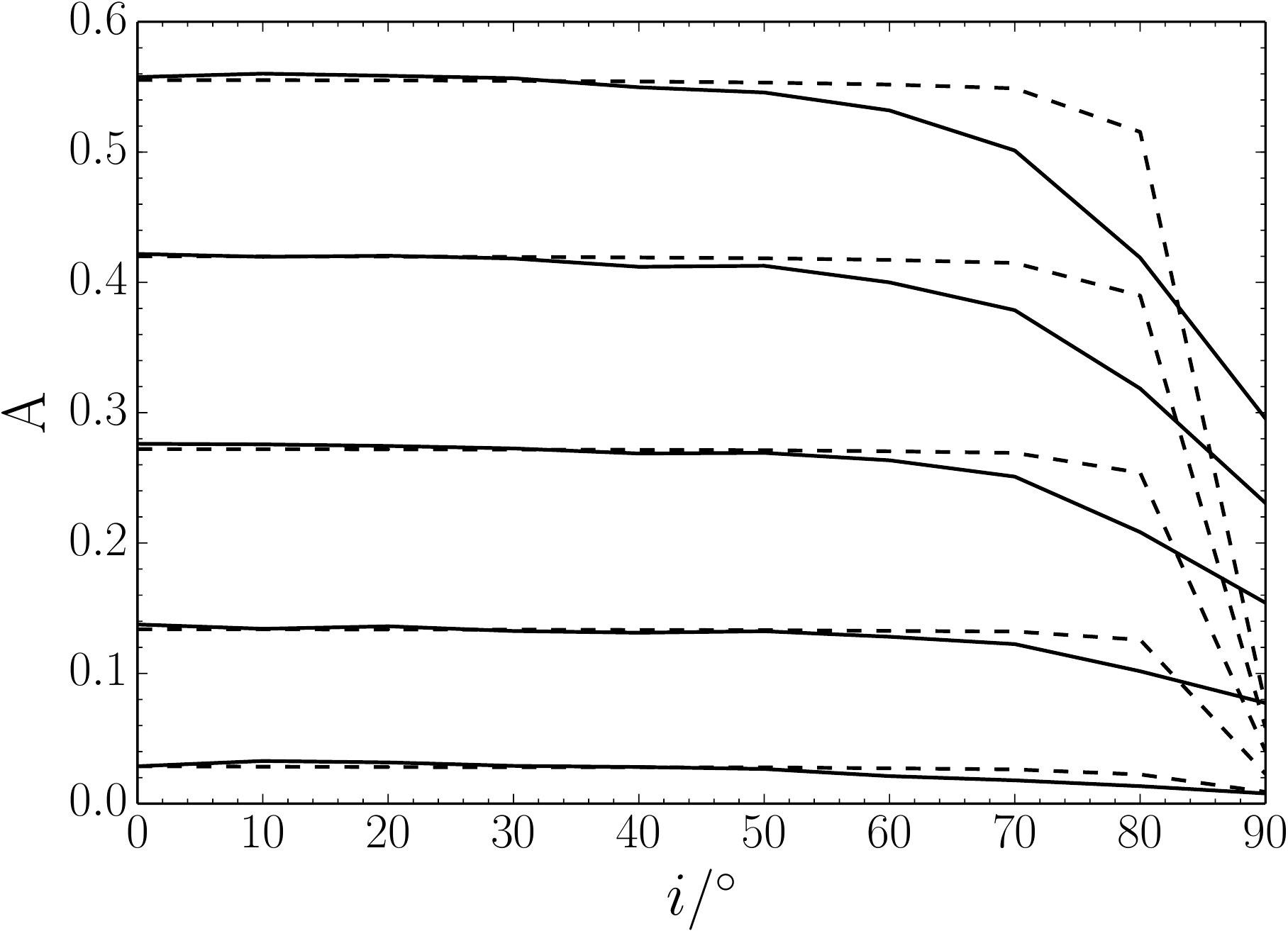}
 \end{center}
 \caption{Comparison of recovered \textit{A} changing with inclination. \newtwo{The solid/dashed line pairs represent models with different levels of intrinsic asymmetry (corresponding to the five models represented in Fig.\ \ref{lop_example}).}
 The dashed lines represent a simple rotation of the galaxy moment zero image around the major axis as presented in \protect\cite{Holwerda2011I}. The solid lines show the trends for \textit{A} in full 3D models which have been set to increasing inclinations. The full 3D model approach already shows a decrease in recovered \textit{A} at inclinations of about 50-60 degrees.
 }
 \label{inclination_3D2D}
\end{figure}

For every galaxy we calculated the Asymmetry parameter from the zero-th moment map. Additionally, we used the moment maps at zero inclination and rotated them around the major axis as described in \cite{Holwerda2011I}. The results are shown in Fig.\,\ref{inclination_3D2D}. The 2D approach does not show a significant change in recovered \textit{A} up to an inclination of 80 degrees. The full 3D approach already shows a decrease of the recovered \textit{A} for lower inclinations. The difference can be explained by the thickness of the disk. The assumption of an infinitely thin disk in the 2D approach does not hold for real galaxies. Inclination limits should thus be derived from full 3D models.

\begin{figure}
 \begin{center}
  \includegraphics[width=0.45\textwidth]{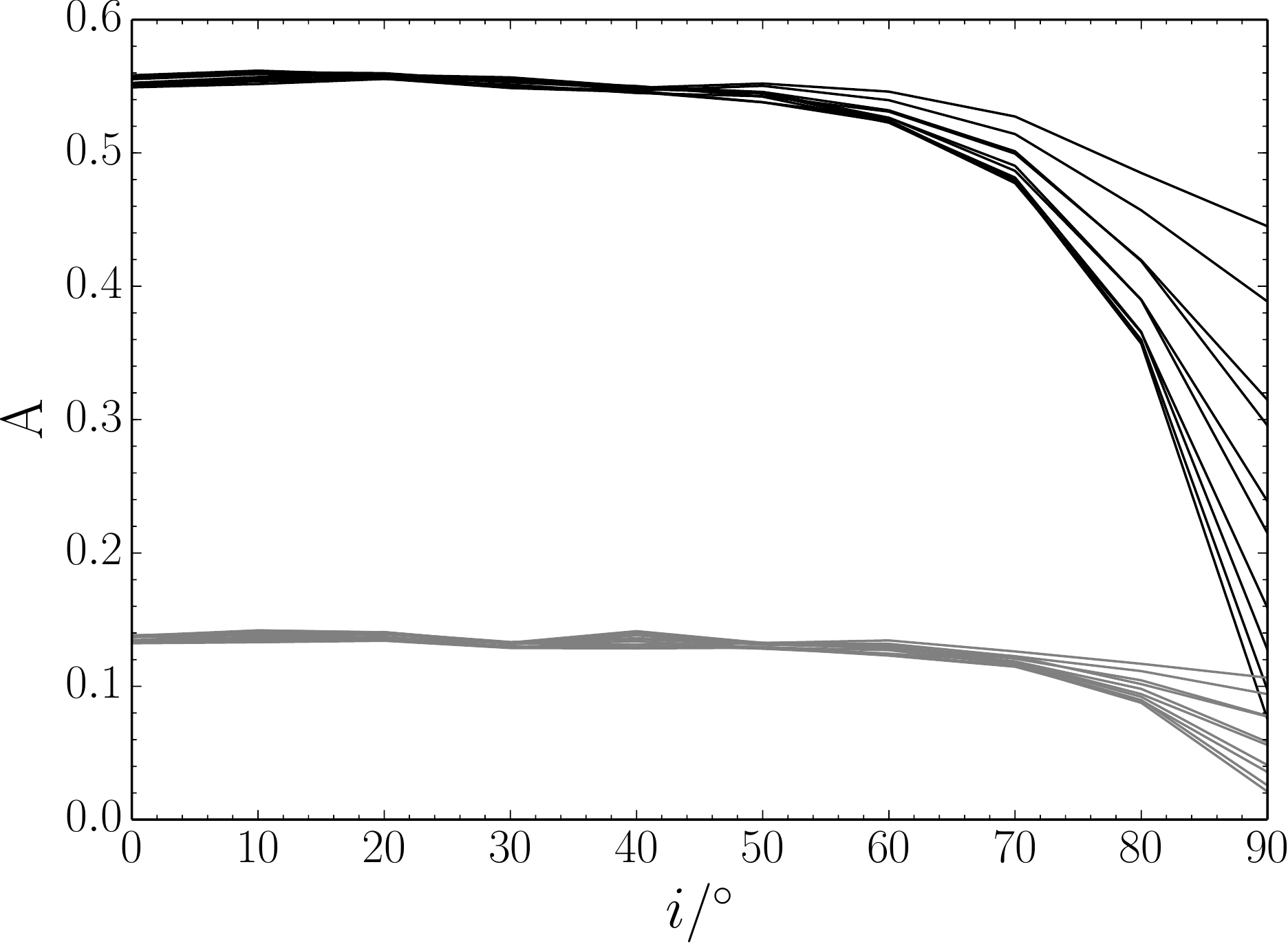}
 \end{center}
 \caption{Noiseless \tir models with increasing inclination and different azimuthal position of lopsidedness in the galaxy disk. \new{The light grey and black lines represent two models with low and high intrinsic morphological lopsidedness (corresponding to the $lop=0.25$ and $lop=1.0$ models in Fig.\ \ref{lop_example}). The different lines of the same grey scale level stem from the same galaxy model. The only difference is the azimuthal position of the source of lopsidedness within the galaxy disk.} The measured A decreases with increasing inclination. The steepness of the decrease depends on the azimuthal position of the lopsidedness in the galaxy. Models with azimuthal positions aligned with the major axis feature a less steep decrease than models with lopsidedness
 aligned with the minor axis.}
 \label{inclination}
\end{figure}

Another important effect that is directly connected to the inclination is the position of the lopsidedness within the galaxy. In our models we changed the azimuthal position of the lopsidedness within the galaxy and calculated the \textit{A} parameter from the zero-th moment map generated from the full 3D cube. In Fig.\,\ref{inclination} we show the results for two models with low (grey) and high (black) degree of lopsidedness. The different curves for each galaxy represent models with lopsidedness at different azimuthal positions between the major and minor axis. The curves that feature a steeper slope at high inclinations represent the models where the lopsidedness is positioned along the minor axis of the galaxy. For models with lopsidedness along the major axis the decrease of the recovered \textit{A} towards high inclinations is more shallow.

In general, one can conclude that the influence of inclination on the \textit{A} estimation already becomes significant at around $50-60^{\circ}$. Therefore, lopsidedness recovered from galaxies at higher inclinations should be seen as a lower limit or excluded from statistical analyses when using this definition for the Asymmetry parameter.

\subsection{Asymmetry is the preferred parameter}\label{section:other_parameters}

Summarising our findings from the last sections, we find that the Asymmetry parameter is the most useful \new{morphometric} parameter for the separation of galaxies with different degrees of lopsidedness. 
\new{We also find a signal to noise limit of $\sim 100$ below which the measured Asymmetry parameter deviates significantly from the intrinsic one.}
It is still possible to separate objects with different degrees of lopsidedness down to a resolution of 3 beams across the major axis. 
Additionally, we find an inclination limit of about $60^{\circ}$ above which the decrease in the measured Asymmetry parameter depends strongly on the azimuthal position of the source of the lopsidedness within the disk.

In a next step we used our models to investigate further how well the parameters can (alone and in combination with each other) distinguish between objects of different morphological lopsidedness. 
We restricted the model galaxies to three versions of lopsidedness: non-lopsided, intermediate and highly lopsided \new{(corresponding to lopsidedness multipliers of 0, 0.5 and 1, see Fig.\ \ref{lop_example})}. Furthermore, we adjusted the properties to the findings from the previous sections and chose inclinations from 0 to 60 degrees, resolutions in beams across the major axis from 3 to 20 and signal to noise levels above 100. We estimated the parameters Asymmetry, Concentration, Gini, $M_{20}$ and \textit{GM} as defined in \cite{Holwerda2011II}. 

\begin{figure*}
 \begin{center}
  \includegraphics[width=0.7\textwidth]{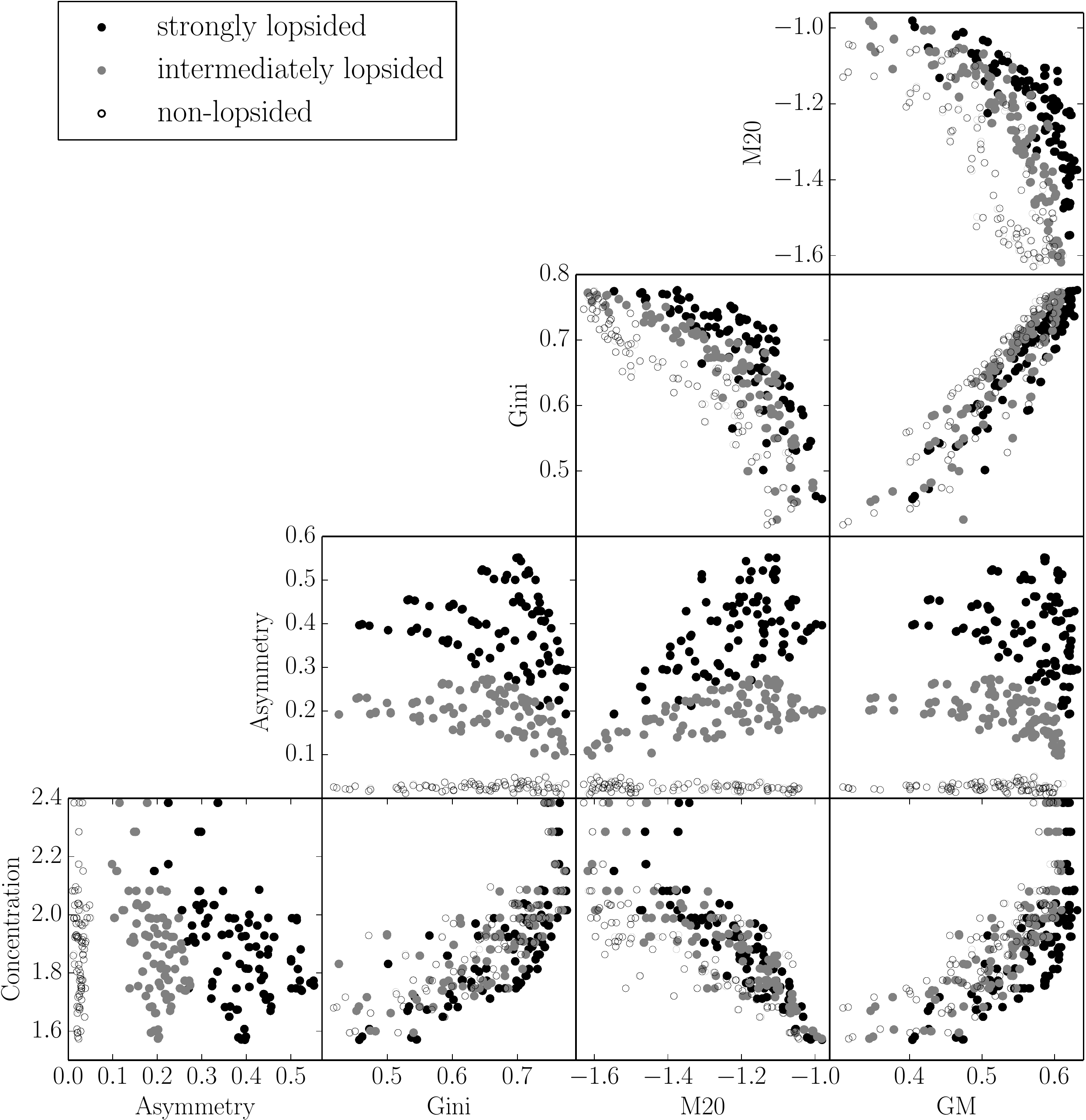}
 \end{center}
 \caption{Correlations between \new{morphometric} parameters measured on \tir model galaxies with varying signal to noise, \HI radius, resolution, and inclination, taking into account the limitations determined in sections \ref{subsubsec:s2n}, \ref{eff:resolution} and \ref{eff:inclination}. \newtwo{We restricted the models to three of the lopsidedness categories illustrated in Fig.\ \ref{lop_example}: strongly lopsided ($lop$=1.0), intermediately lopsided ($lop$=0.5) and non-lopsided ($lop$=0).}
 The Asymmetry parameter in combination with all other parameters clearly separates lopsided from non-lopsided galaxies, whereas there are slight to strong overlaps between the remaining parameters.}
 \label{corr_parameters}
\end{figure*}

The results are shown in Fig.\ \ref{corr_parameters}. Therein, all panels showing the combination of parameters which include the Asymmetry parameter show a pronounced separation between lopsided and non-lopsided galaxies. In contrast, the strongest overlaps occur in combinations with the Concentration parameter and in the Gini-\textit{GM} combination. The $M_{20}$ parameter also shows overlaps with other parameters. In the combinations with Gini and \textit{GM} these overlaps still allow for a separation between certain fractions of the non-lopsided and strongly lopsided objects.

\begin{figure}
 \begin{center}
  \includegraphics[width=0.45\textwidth]{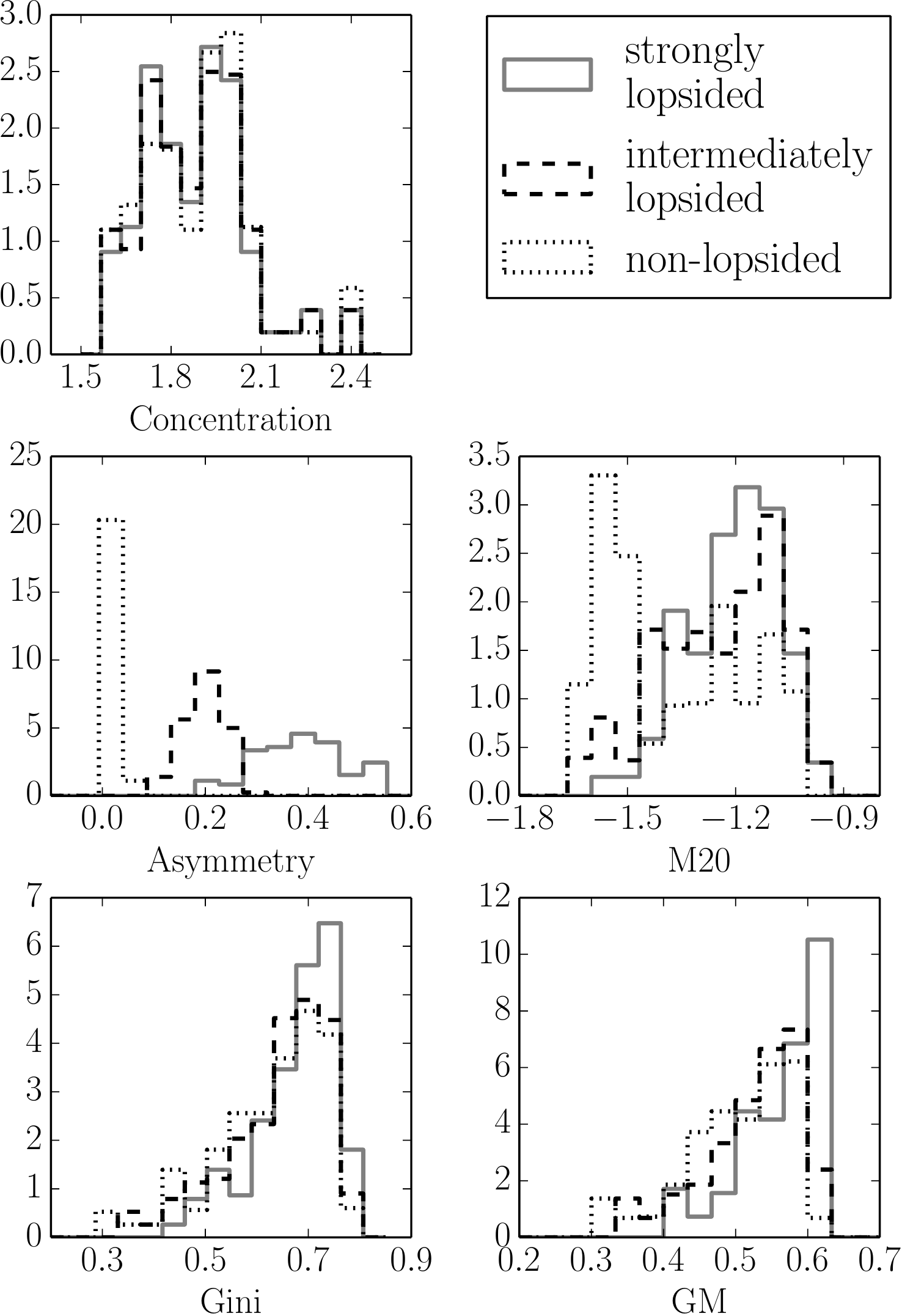}
 \end{center}
 \caption{Morphological parameters measured on \tir model galaxies with varying signal to noise, \HI radius, resolution, and inclination, taking into account the limitations determined in sections \ref{subsubsec:s2n}, \ref{eff:resolution} and \ref{eff:inclination}. \newtwo{We restricted the models to three of the lopsidedness categories illustrated in Fig.\ \ref{lop_example}: strongly lopsided ($lop$=1.0), intermediately lopsided ($lop$=0.5) and non-lopsided ($lop$=0).} The Asymmetry parameter clearly separates lopsided from non-lopsided galaxies.}
 \label{hist_parameters}
\end{figure}

Fig.\ \ref{hist_parameters} shows the separation of the different lopsidedness groups when only single parameters are used. Obviously, in the presence of high signal to noise, the \textit{A} parameter is the most promising one with regards to the separation between galaxies with different degrees of lopsidedness. For all other parameters, every group with a particular degree of lopsidedness spans the complete range of possible values.

Considering the good separation of the three different model groups it is reasonable to prefer the Asymmetry parameter to identify galaxies with lopsided distributions, if the resolution and inclination of the objects are within the determined limits and the \textit{A} parameter can be corrected for the noise bias.

\label{subsection_asymm_dependence}

\section{Correction of Asymmetry values at low S/N}\label{section:bias}
    The \textit{A} parameter turns out to be the most suitable parameter for the evaluation of the degree of lopsidedness within the total \HI image of a galaxy. It shows the smoothest and least steep trends in the dependence on the different data qualities or galaxy properties. Furthermore, in the presence of high signal to noise it can be used independent from the other parameters to separate lopsided from non-lopsided objects.
In this section we investigate whether its measured value can be corrected to take into account the effect of various data characteristics.

The effects of inclination and resolution cannot be corrected for because both are due to a loss of information in the data. However, these effects can be taken into account by considering the measured \textit{A} as a lower limit of the intrinsic \textit{A} \new{(the underlying Asymmetry value without noise)}. Galaxies within groups of the same resolution and inclination could still be evaluated with respect to each other. In contrast, the effect of noise is additive and in combination with the aforementioned effects makes an interpretation of recovered \textit{A} values complicated if not impossible.

We define the positive offset, the noise bias, which becomes significant for galaxies with mean signal to noise lower than 100, as follows:

\begin{equation}
 A_{\textrm{measured}} = A_{\textrm{intrinsic}}+\textrm{Bias}.
 \label{eqn_bias1}
\end{equation}

A very simple approach to avoid the noise bias is to increase the signal to noise in the moment maps. This can be achieved by e.g. smoothing or de-noising in Fourier space. However, this can either lead to a loss of information or to an introduction of new patterns which are not real. CBJ00 presented an approach for the prediction of the noise bias. Therein, the mask defining the area of the galaxy is shifted around in areas outside the mask to estimate the numerator of the asymmetry only in regions which contain noise (and background). The corrected \textit{A} parameter is definded as:

\begin{equation}
\label{eqn_conselice_bias}
 A_{\textrm{corrected,CBJ00}} = \frac{\sum{\left|I-I_{180}\right|}}{2{\sum{\left|{I}\right|}}}-\frac{\sum{\left|B-B_{180}\right|}}{2\sum{\left|{I}\right|}}, 
\end{equation}

\noindent where B has the same shape as I (using the same mask), but defines a region outside of the galaxy that contains background as well as noise. The minuend on the right hand side of Eq.\ \ref{eqn_conselice_bias} corresponds to the Bias in Eq.\ \ref{eqn_bias1}. Taking into account that \HI data ideally has a zero background level, this approach would in general lead to an overcorrection for a non-zero noise level. To illustrate this point, consider the following:

\begin{align}
&\sum{\left|I-I_{180}\right|}-\sum{\left|B-B_{180}\right|}\nonumber\\
&=\sum\left|(I_0+n)-(I_{0,180}+n_{180})\right|-\sum\left|n-n_{180}\right|\nonumber\\
&\leq\sum\left|I_0-I_{0,180}\right|
\end{align}

where $I_{0(,180)}$ are the pixel intensities of the image without noise and $n_{(,180)}$ samples of Gaussian noise (and their respective 180 degree rotated versions); the sum of which constitute the components of the moment map ($I = I_0+n$). It becomes apparent that the bias corrected Asymmetry parameter always represents a lower limit for the intrinsic \textit{A}:
\begin{align}
 \Rightarrow A_{\textrm{intrinsic}}  \geq A_{\textrm{measured}} -  \textrm{Bias(CBJ00)} = A_{\textrm{corrected,CBJ00}}                              
\end{align}
Moreover, it should be noted that this approach does not take into account the fact that the bias has a larger influence on pixels with low intensity. It is therefore not suitable as a correction for the noise bias in \HI data.

A new approach that we propose here is to use model galaxies to approximate the noise bias. To investigate the feasibility of this approach, we built a data base of \tir models (set S1) changing different properties such as the noise level (hence the signal to noise), the size of the galaxy
relative to the angular resolution, the inclination, the intensity and azimuthal position of the morphological lopsidedness, and the \HI radius. We again use the surface brightness definition from \cite{Martinsson2011} and the rotation curve displayed in Fig.\ \ref{tirific_params}. The lopsidedness is introduced as a first order harmonic surface brightness distortion. \new{We used \SoFiA to determine the pixels that contain emission. This time we used the cubes including noise, with the result that some galaxies could not be found due to low S/N}. For our models we estimated not only the noise bias, but also values that could be determined from observational data, e.g. the mean signal to noise, the total flux, the \textit{A} value (which includes the bias) and the number of 2D and 3D pixels in the mask, which defines the region of the galaxy. As a simple test, we created two additional sets of galaxies, where we varied the input galaxy properties (e.g. inclination, resolution, etc.) in the same range covered by the objects of set S1. The lopsidedness in the first set (S2) was generated by a first order harmonic surface-brightness distortion while in the second set (S3) lopsidedness was generated by a third order surface-brightness distortion \new{(see Fig.\ \ref{example_order} for an example). 
\begin{figure}
 \begin{center}
  \includegraphics[width=0.40\textwidth]{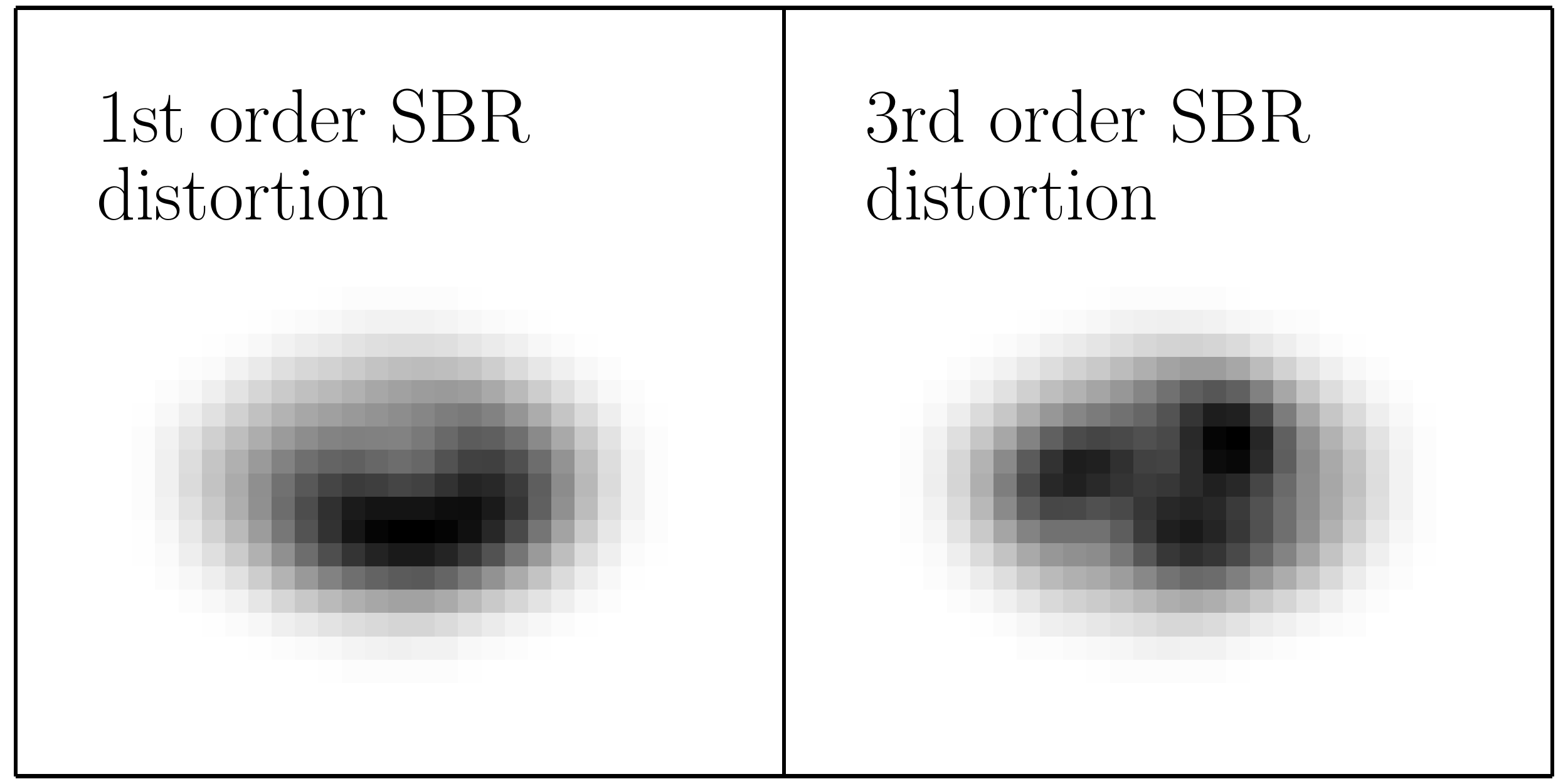}
 \end{center}
 \caption{Example galaxies with lopsidedness introduced by a first order (left panel) and a third order (right panel) harmonic surface brightness distortion.}
 \label{example_order}
\end{figure}
}

The different parameters of the model sets are listed in Table \ref{table:input}.

\begin{table}
  \caption{Model parameters for the different galaxy model sets}
  \begin{tabular}{p{2.2cm}|p{2.2cm}|p{1.1cm}|p{1.1cm}}
    Parameters                       & Galaxy set S1                                              & Galaxy set S2 & Galaxy set S3\\ \hline
    Inclinations [$^{\circ}$]                    & \parbox[t]{5cm}{0,10,20,...,90}   & \multicolumn{2}{c}{15, 35, 55} \\ \hline
    Azimuthal angle of asymmetry [$^{\circ}$]    & \parbox[t]{5cm}{0,10,20,...,90}   & \multicolumn{2}{c}{15, 45}    \\ \hline
    Lopsidedness multiplier             & 0, 0.25, 0.5, 0.75, 1.0                                   & \multicolumn{2}{c}{0.15, 0.45, 0.75} \\ \hline
    Order of harmonic SBR distortion & 1st                                                       & 1st           & 3rd\\ \hline
    $R_{\textrm{\HI}}$ in rings      & 1, 3, 5, 7, 9, 11                                         & \multicolumn{2}{c}{4, 8, 10}\\ \hline
    $d_{\textrm{galaxy}}$ [\# beams] & 20, 15, 10, 7, 5, 3, 1                                    & \multicolumn{2}{c}{12, 8, 6, 4}\\ \hline
    Noise amplitude multiplier       & \parbox[t]{2.2cm}{0,5E-8,5E-7,\\5E-6,5E-5,\\1.58E-5,5E-4,\\5E-3,5E-2,\\5E-1,1.58E-1,\\1.58,5,50}& \multicolumn{2}{c}{\parbox[t]{2.2cm}{3E-4,3E-3,3E-2, 3E-1,3,30}}\\
  \end{tabular}
  
  \label{table:input}
\end{table}

\begin{figure}
 \begin{center}
  \includegraphics[width=0.40\textwidth]{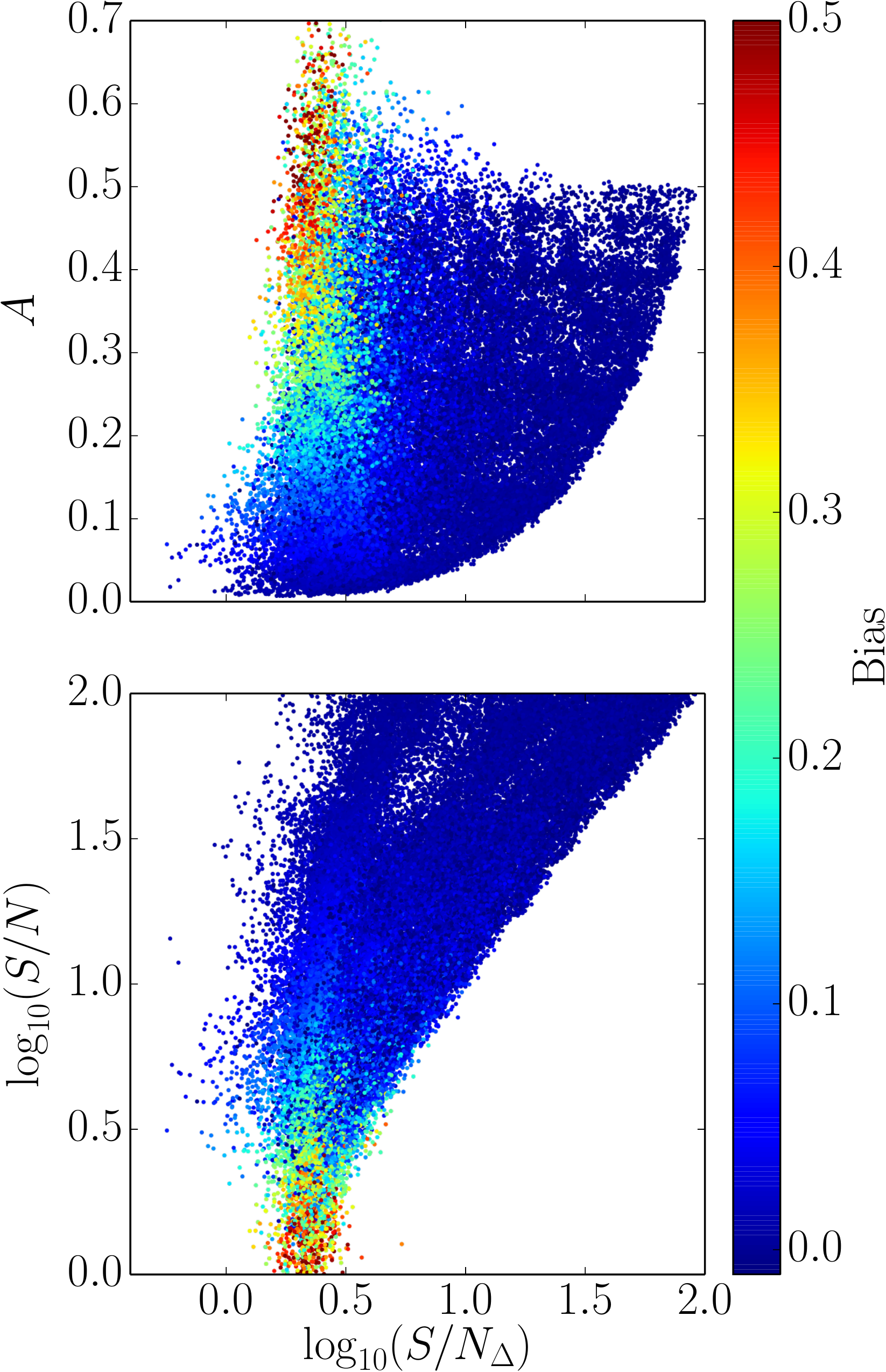}
 \end{center}
 \caption{Bias dependence of the galaxies in the model data base S1 on the \new{morphometric} \textit{A} parameter, the mean S/N and the $S/N_{\Delta}$ parameter. We find a smooth trend with higher bias values concentrating in regions of high \textit{A} values, low signal to noise and low $S/N_{\Delta}$. The smooth trend makes these three parameters good candidates for a noise bias correction.}
 \label{s2ndiff_a2d}
\end{figure}

Since we know that the bias is significant in particular for low pixel differences between the \HI image and its rotated version we define another parameter, which measures the signal to noise of the intensity difference $S/N_{\Delta}$:
\begin{equation}
 S/N_{\Delta} = \left\langle \frac{|I(i,j)-I_{180}(i,j)|}{\sqrt{\frac{1}{2}(N(i,j)+N_{180}(i,j))}\sigma} \right\rangle.
 \label{eq:s2ndiff}
\end{equation}
$I$ and $I_{180}$ are the pixel intensities, $N$ and $N_{(180)}$ the number of channels along the line of sight through the cube that were added up to the pixel value $I$ and $I_{180}$ respectively, and $\sigma$ is the RMS noise in the cube. If we plot the bias for the model galaxies of set S1 as a function of the measured \textit{A} parameter, the mean signal to noise and the newly defined $S/N_{\Delta}$ we find a smooth trend (see Fig.\ \ref{s2ndiff_a2d}), which suggests that these three parameters are well suited for estimating the bias in galaxy \textit{A} measurements.

\begin{figure*}
 \begin{center}
  \includegraphics[width=0.60\textwidth]{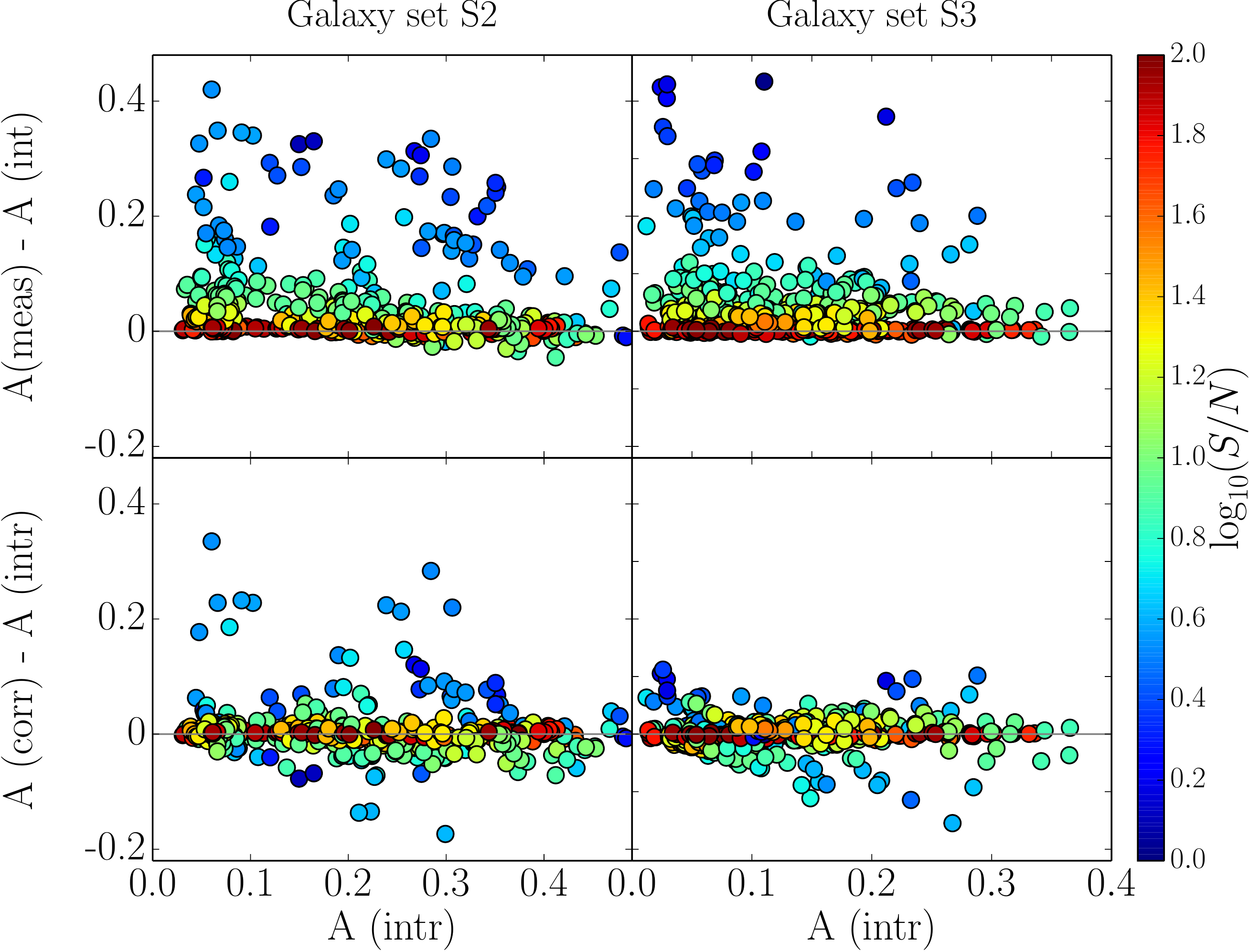}
 \end{center}
 \caption{Intrinsic, measured and corrected \textit{A} values for the galaxy model sets with first order (S2) and third order (S3) harmonic surface brightness distortions. Both corrections result in a better agreement between the intrinsic and the recovered \textit{A} values. 
 }
 \label{A_bias_corrected}
\end{figure*}

We estimated the \textit{A} parameters for all the galaxies in our training (S1) and test sets (S2,S3) with and without noise. The results for the two test sets can be seen in the upper panels of Fig.\ \ref{A_bias_corrected}.
We plotted the \new{difference between the measured and the} intrinsic \textit{A} parameter against the measured one. The measured \textit{A} is clearly influenced by the noise bias\new{, hence the positive offset}. The colours of the points are indicators of each galaxy's signal to noise, which is applied on a logarithmic scale. We only plot galaxies that have a S/N below 100, since \textit{A} values do not need correction for S/N values above 100. As expected, the models with lower mean signal to noise feature a higher bias value.

In a next step we used the machine learning library {\sc scikit-learn}\footnote{http://scikit-learn.org} and the first set of galaxy models to interpolate the noise bias for the galaxies in the test sets. For the bias interpolation we used a 3D parameter space, which consists of the new parameter $S/N_{\Delta}$ as defined in Eq.\ (\ref{eq:s2ndiff}), the mean signal to noise (see Eq.\ \ref{eq_s2n1} and \ref{eq_s2n2}) and the measured \textit{A} parameter of the galaxies including noise. These parameters can be measured on a pixel base and do not require any further knowledge about the galaxy structure. Since our model set S1 is large enough to slow down any algorithm significantly, we divide the parameter space into subspaces and correct the test sets by interpolating in these smaller parameter regions. This speeds up the process notably because some parameter regions do not have to be explored due to the absence of test examples in these regions. Furthermore, we excluded all test examples that turned out to have a negative total flux assuming that these sources would not have been found using the source finder.

We use the machine learning suit {\sc scikit-learn} with the galaxy set S1 as the training set. The trained algorithm is
then used to predict the bias values for the test sets S2 and S3, which are subtracted from the estimated \textit{A} parameter for the galaxies with noise. The \new{difference between the} corrected \textit{A} value \new{and the intrinsic one} is shown in the lower panels of Fig.\ \ref{A_bias_corrected}. The correction works quite well for both data sets. \new{Offsets from the intrinsic values spread symmetrically around zero.}
This is particularly surprising since the model set S1 and test set S3 have been generated using different forms of lopsidedness.

\begin{figure*}
 \begin{center}
  \includegraphics[width=0.60\textwidth]{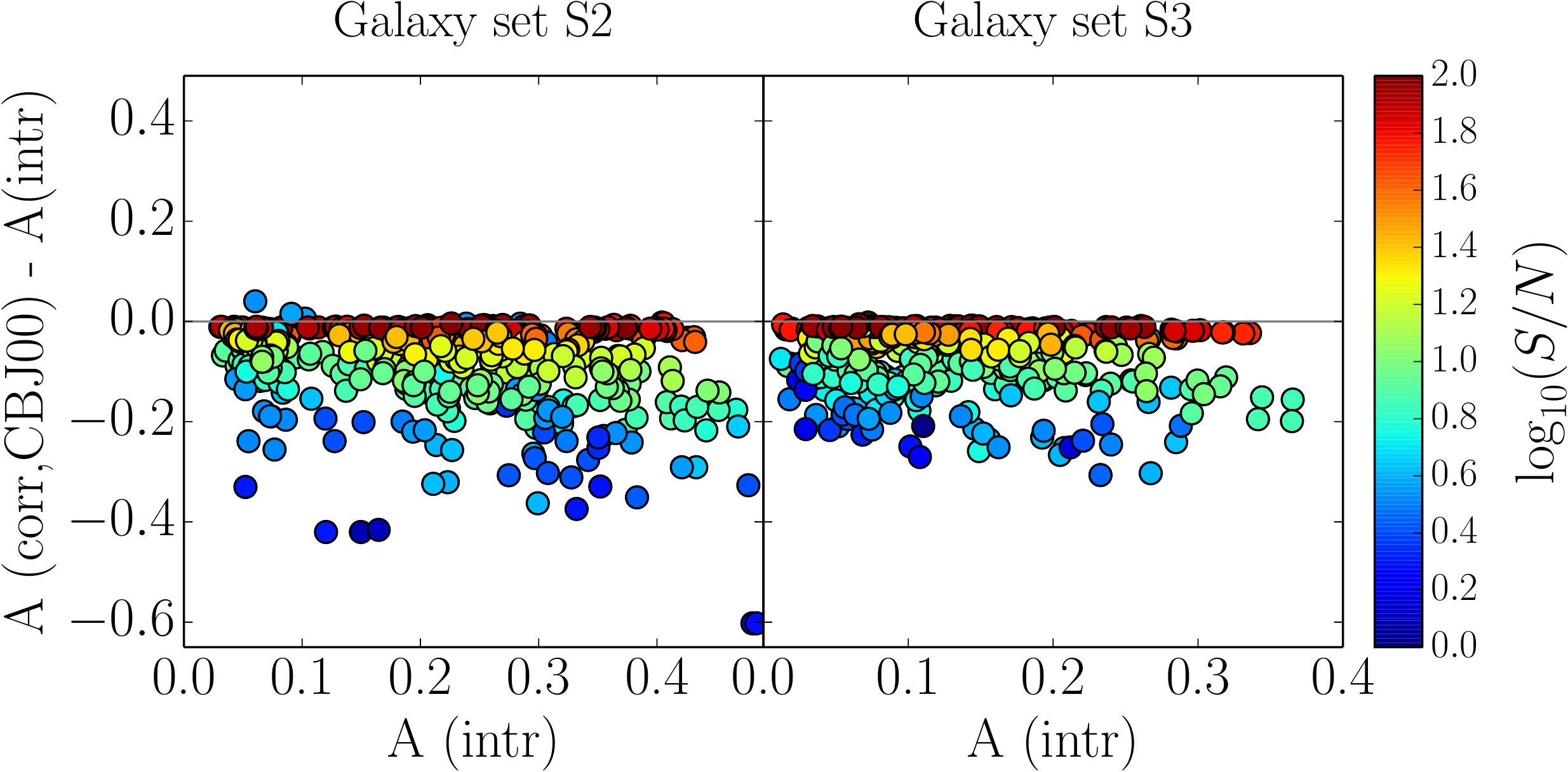}
 \end{center}
 \caption{Corrected \textit{A} values for the galaxy model sets with first order (S2) and third order (S3) harmonic surface brightness distortions using the bias correction from \protect\cite{Conselice2000}. As expected the Asymmetry parameter values are over-corrected.}
 \label{A_bias_corrected_conselice}
\end{figure*}

For a comparison we also estimated the bias using the prescription given in CBJ00. We shifted the masks around the galaxies only varying the spatial coordinates. We chose shift positions that span a 3 by 3 grid with a grid spacing equal to the galaxy size in RA and Dec and the actual galaxy position being in the central grid point. This way we end up with eight bias measurements for a galaxy. The final bias is estimated as the median of this bias sample distribution. The results are shown in Fig.\ \ref{A_bias_corrected_conselice}.

As expected, the resulting Asymmetry parameter values are over-corrected. But as mentioned earlier, this particular correction method is not suited for \HI data.

\begin{figure*}
 \begin{center}
  Galaxy set S2 \hspace{6cm} Galaxy set S3\hspace{1cm}\   
  
  \includegraphics[height=0.231\textheight]{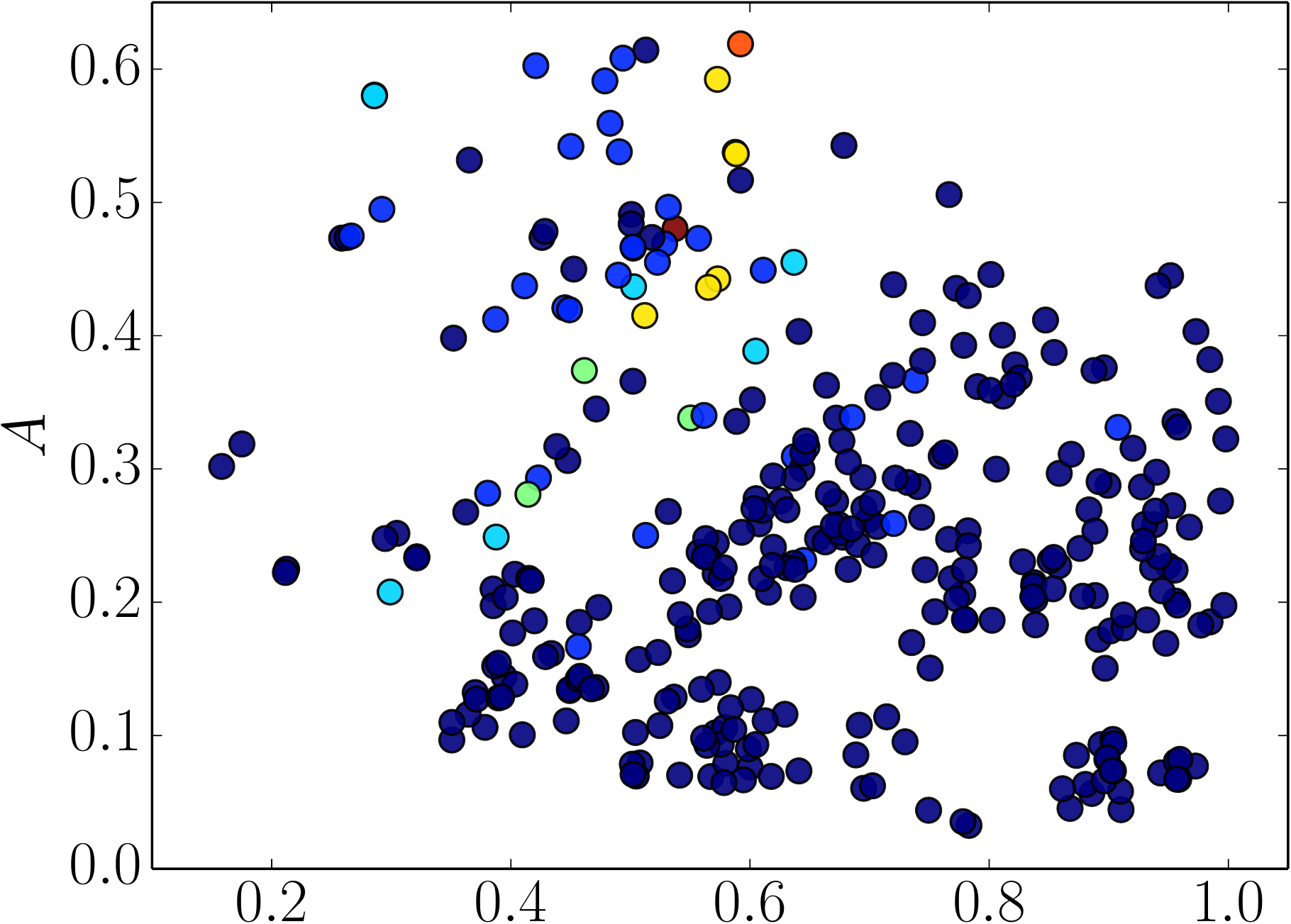}
  \hspace{0.4cm}
  \includegraphics[height=0.231\textheight]{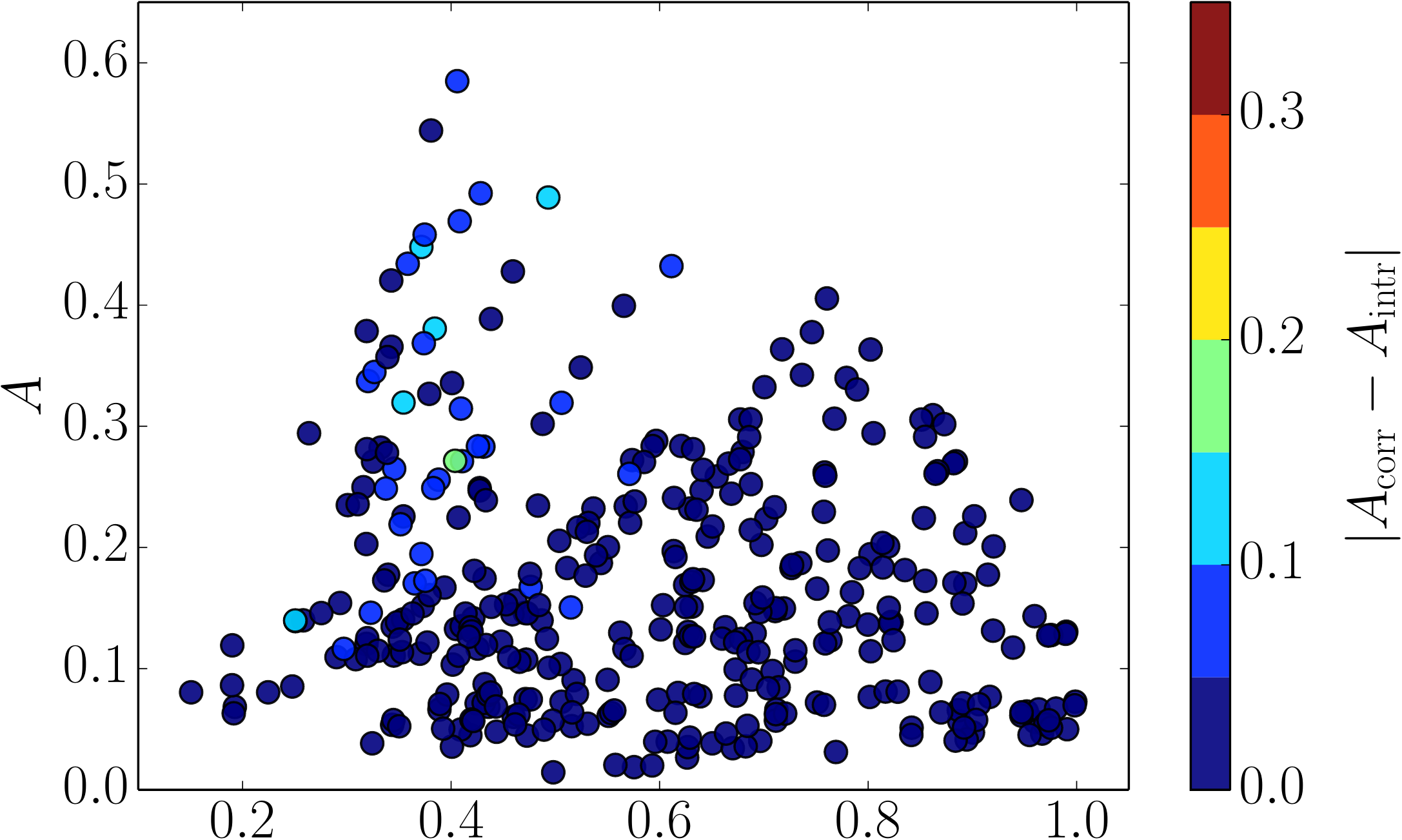}
  
  \includegraphics[height=0.231\textheight]{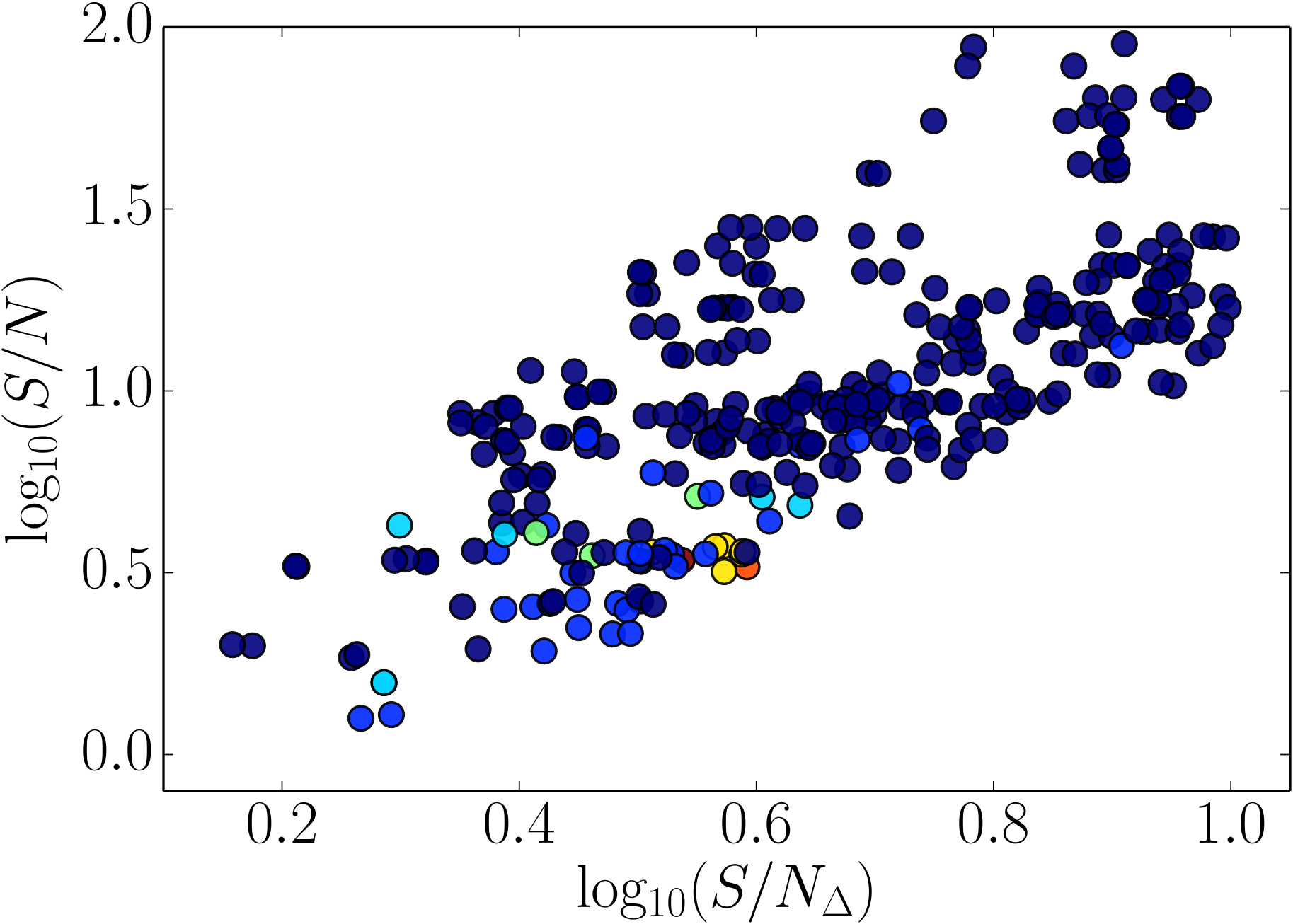}
  \hspace{0.4cm}
  \includegraphics[height=0.231\textheight]{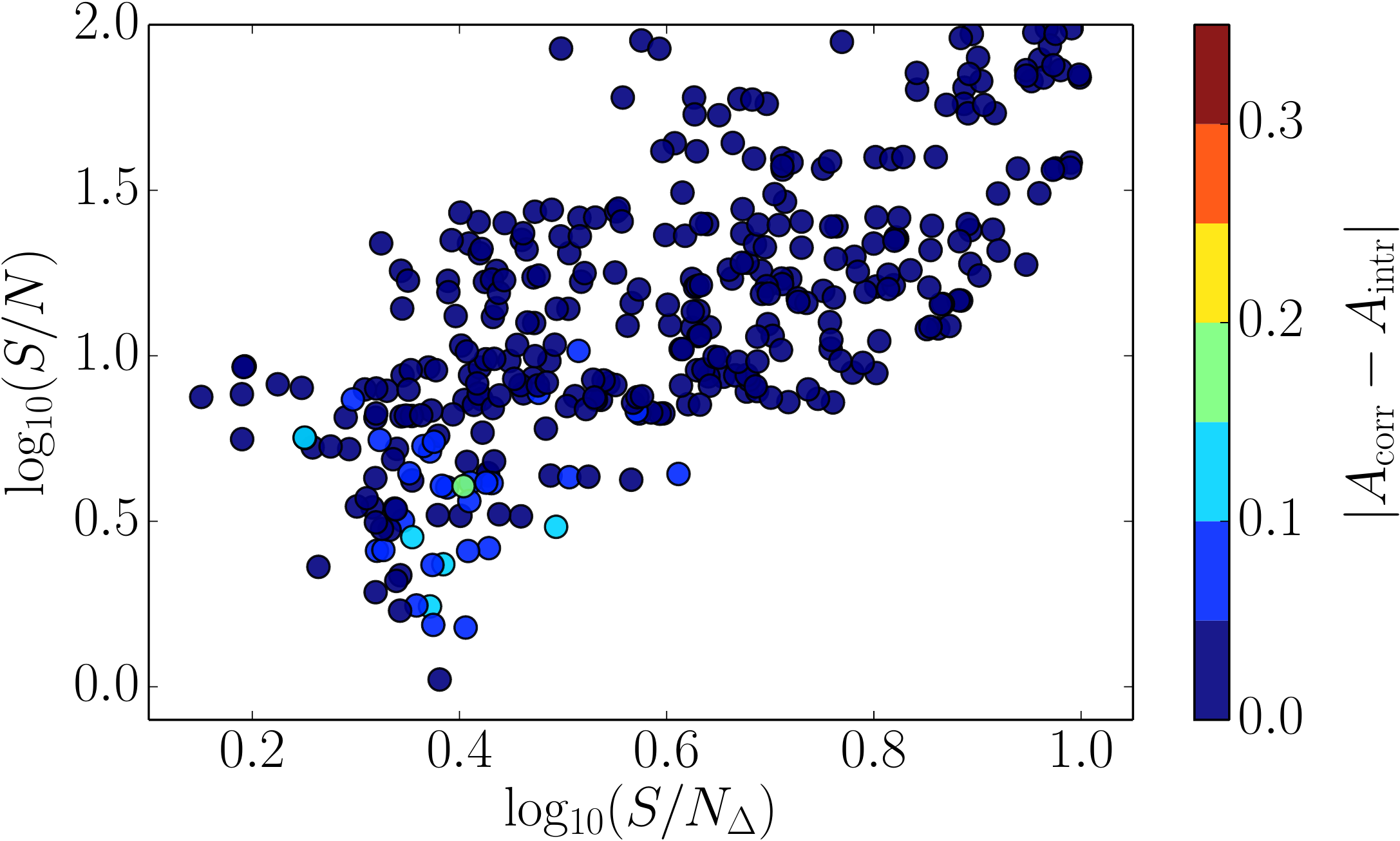}
  
 \end{center}
 \caption{Quality evaluation of the \textit{A} parameter correction with respect to the position in parameter space. 
 The colours indicate to what extent the measured value can be corrected within the original value.
 Both test sets show sufficient correction within 0.35, the majority of objects can be corrected within 0.1 of the original value. There is no trend for the correction quality in the 3D parameter space used for correction.
 }
 \label{fig:corr_quality_full}
\end{figure*}

\new{For the model based correction method we further investigated how well the \textit{A} parameter can be bias corrected within the parameter space that was used for correction.
We show the results in Fig.\ \ref{fig:corr_quality_full} using the same two projections as in Fig.\ \ref{s2ndiff_a2d}.
We find that there is no clear trend in the correction quality in the shown parameter space. \new{Most galaxies can be corrected within 0.05 of the original value. There are a few outliers in the galaxy set S2 which have higher deviations with values for the difference between intrinsic and estimated \textit{A} up to $0.33$.}
}

These results lead us to the conclusion that the presented method is suitable for e.g. a simple classification into non-lopsided, intermediately lopsided and strongly lopsided objects, as it is indeed possible to correct the noise bias to a certain extent in a very simple way without the need for additional knowledge about any properties of the galaxy.

\section{Conclusions and outlook}\label{section:conclusion}
    We found that the application of the optical parameters Concentration, Asymmetry, Gini, M$_{20}$ and \newtwo{GM} adopted and defined in \cite{Holwerda2011II} on \HI data does not produce meaningful results unless effects like the signal to noise of the data as well as the inclination and resolution of the galaxies are taken into account. We found that above a signal to noise limit of about 100, the optical parameters can be used on \HI data without correction. For a mean signal to noise in the \HI maps below this threshold, it is necessary to apply a correction.

We have found the Asymmetry parameter to be the most useful among the set of investigated parameters for measuring galaxy lopsidedness. For low signal to noise the Asymmetry parameter features a noise bias, which is always positive. Inclination and resolution effects influence the result in a very similar way. The intrinsic \textit{A} can be recovered to a certain degree down to a resolution of about 3 beams across the major axis. Since the resolution as well as the inclination effects cause a loss of detail, it is not possible to recover the full information about the lopsidedness of a galaxy. The recovered \textit{A} parameter should always be seen as a lower limit. When comparing galaxy samples they should therefore be sorted into categories of similar resolutions and inclinations. We found an inclination limit of approximately 60$^{\circ}$ above which the steepness of the decrease of \textit{A} depends strongly on the azimuthal position of the lopsidedness within the disk.
For lopsidedness aligned with the minor axis, this decrease is steeper than for the ones aligned with the major axis.

For the correction of the effects of low S/N ($\ll 100$) we have investigated the approach of using a galaxy model data base to find the noise bias as a function of S/N and the recovered Asymmetry parameter. We find that it is possible to correct the noise bias, provided that the galaxy models in the correction set are similar to the galaxies that need bias correction. Our models are all similar in the shape of the surface brightness profile and the rotation curve. The model data base should be updated with different versions of surface brightness profiles and rotation curves. One could consider building a data base using the properties determined from data of real objects that have been classified visually. Another approach that might be considered is the use of real galaxies with high signal to noise. These could be used as templates to investigate the effects of resolution and signal to noise. However, since only a small number of observed galaxies that have been observed are in the signal to noise regime necessary ($\approx$ 100), they can only be seen as an addition to models if they are used for bias correction.
The machine learning tool {\sc scikit-learn} is a very useful tool to interpolate the bias for objects using a pre-built data base. Since it is mostly only necessary to classify different categories of asymmetry or lopsidedness (e.g. strong, intermediate, no lopsidedness), one could consider using a discrete classification scheme, which would eventually also offer more options to employ more sophisticated machine learning techniques, such as neural networks to classify galaxies.

In a next step the methods should be tested on real data. An ideal test case is the WHISP survey galaxy sample. A significant number of galaxies has been \new{classified visually} \citep{Swaters2002,Noordermeer2005} and can be used as a test set for the verification of the method. \newtwo{This is the subject of a subsequent study.} A scientific example for the application of the Asymmetry parameter is the investigation of the relation between asymmetry or morphology and environment density of galaxies using surveys such as WHISP or Atlas$^{3D}$, which contain galaxies in different environments. This will be the subject of a forthcoming paper.

Eventually, the asymmetry parameter should be extended to 3D which would allow for the use of the full information contained in a data cube. In contrast to detailed 3D modelling (e.g. with \tir), this approach would represent an alternative for the estimation of a 3D asymmetry for objects with low angular resolution.

\section*{Acknowledgments}
NG and JMvdH acknowledge support from the European Research Council under the European Union's Seventh Framework Programme (FP/2007-2013) / ERC Grant Agreement nr. 291531.

We would like to thank the referee for his/her detailed and useful comments and suggestions.

\appendix

\section{Appendix}

\subsection{\tir input}

\begin{figure*}
 \begin{center}
  \includegraphics[width=0.7\textwidth]{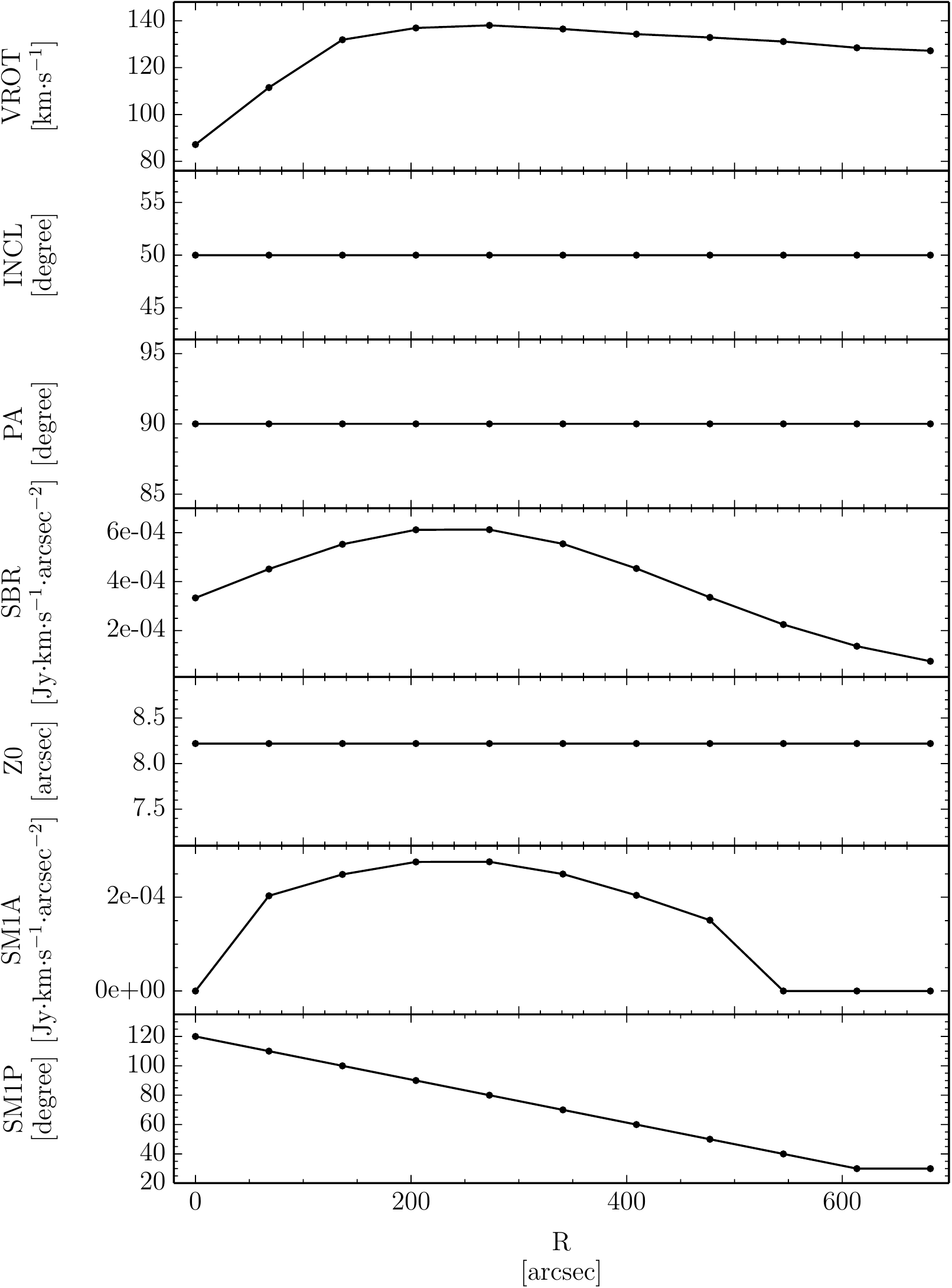}
 \end{center}
 \caption{Example model input parameters for \tir. Models consist of 11 rings. Every ring is characterised by a rotational velocity (VROT), an inclination (INCL), a position angle (PA), a surface brightness (SBR), a scale height(Z0) and the amplitude (SM1A) and phase (SM1P) of the first order harmonic surface brightness distortion.}
 \label{tirific_params}
\end{figure*}

Fig.\ \ref{tirific_params} shows some example \tir input parameters for the galaxy models displayed in Fig.\ \ref{s2n_example}. The model consists of 11 rings. The input parameters include the rotation curve, the inclination and position angle, the surface brightness profile, the scale height and the parameters that regulate the lopsidedness. For the plot in Fig.\ \ref{s2n_example} the model was re-gridded to a resolution of 7 beams across the major axis.

\bsp

\label{lastpage}

\end{document}